\documentclass[12pt]{article}
\usepackage{amsmath}
\usepackage{amssymb}
\usepackage{graphicx}
\usepackage{epsfig}
\usepackage{color}
\usepackage{fullpage}
\usepackage{hyperref}

\newcommand\pubdate{}

\def\STANKI{Dept. of Physics\\
 Massachusetts Institute of Technology , Cambridge, MA 02139}

\def\Title#1{\begin{center} {\Large #1 } \end{center}}
\def\Author#1{\begin{center}{ \sc #1} \end{center}}
\def\Address#1{\begin{center}{ \it #1} \end{center}}

\newcommand\pubblock{\rightline{\begin{tabular}{l}\\
         \pubdate \end{tabular}}}

\newcommand{\bp}{\bar{\varphi}}

\newcommand{\mpl}{m_{\textrm{pl}}}

\newcommand{\tk}{\tilde{k}}
\newcommand{\nl}{\textrm{nl}}
\newcommand{\p}{\textrm{p}}
\newcommand{\z}{(\lambda/g)}
\newcommand{\sbp}{\bar{\phi}}

\date{ }
\begin{document}

\begin{titlepage}
\pubblock

\vfill \Title{Inflaton fragmentation: \\Emergence of pseudo-stable inflaton lumps (oscillons) after inflation.}
\vfill \Author{Mustafa A. Amin\footnote{email: mamin@mit.edu}} \Address{\STANKI}

\begin{abstract}
We investigate the emergence of  large, localized, pseudo-stable configurations (oscillons) from inflaton fragmentation at the end of inflation. We predict the number density of large oscillons, and the conditions necessary for their emergence in a class of inflationary models. Analytic estimates are provided for a $3+1$ and $1+1$-dimensional universe. We test our predictions with detailed numerical simulations in $1+1$-dimensions. We see a zoo of oscillons emerging from the simulations, including the usual small amplitude ``sech" oscillons as well as large ``flat-topped" oscillons. The emergent oscillons account for $\sim 80$ per cent of the energy density of the inflaton. 
\end{abstract}

\vfill
\end{titlepage}
\def\thefootnote{\fnsymbol{footnote}}
\setcounter{footnote}{0}

\section{Introduction}
Inflation \cite{Guth:1980zm, Linde:1981mu, Albrecht:1982wi} is a wonderful mechanism for generating large scale density fluctuations  that, under the influence of gravity, eventually result in the formation of structure in the late universe. However, inflation must eventually end to give rise to a hot, radiation dominated universe consistent  with the success of Big Bang Nucleosynthesis (BBN) \cite{Alpher:1948,Wagoner:1966pv}. Hence, we require the inflaton to eventually decay into Standard Model fields, possibly via intermediaries \cite{Zeldovich:1975,Abbott:1982hn,Dolgov:1982th,Albrecht:1982mp}.

Before the universe thermalizes, the inflaton often undergoes complex spatio-temoporal dynamics. In many cases it fragments on time scales $t\ll H^{-1}$, leading to a turbulent, incoherent state of scalar waves (see for example \cite{Traschen:1990sw,Kofman:1994rk,Shtanov:1994ce,Kofman:1997yn, Podolsky:2005bw,GarciaBellido:2002aj, Felder:2001kt, Felder:2000hj, Tkachev:1998dc, Khlebnikov:1996mc,Greene:1998pb,Dufaux:2006ee,Frolov:2008hy} and for recent reviews and further references see \cite{Allahverdi:2010xz,Frolov:2010sz}).  In this paper, we discuss a class of real-valued, single field models, where the fragmentation can lead to copious formations of remarkably long-lived ($t\gg H^{-1}$), localized, pseudo-solitonic configurations called oscillons. 

Oscillons have been known to exist in non-linear scalar field theories for some time. The earliest investigations can be found in \cite{Bogolyubsky:1976yu,Gleiser:1993pt, Copeland:1995fq} (also see \cite{Honda:2001xg, Adib:2002ff, Farhi:2005rz, Gleiser:2006te, Graham:2006xs, Hindmarsh:2006ur, Fodor:2006zs, Saffin:2006yk, Graham:2006vy, Arodz:2007jh, Hindmarsh:2007jb, Fodor:2008es, Gleiser:2008ty, Fodor:2008du, Fodor:2009xw, Gleiser:2009ys, Gleiser:2007te, Amin:2010jq,  Hertzberg:2010yz}). They are localized in space and oscillatory in time.\footnote{Hence they are not constrained by Derrick's theorem \cite{Derrick:1964ww} which only applies to static configurations}
They are long lived, surviving for many thousands of oscillations. They are quasi-stable, losing energy very slowly through outgoing scalar radiation (for example, see \cite{Segur:1987mg,Fodor:2009kf,Gleiser:2009ys,Hertzberg:2010yz,Farhi:2007wj}).  Unlike $Q$-balls \cite{Coleman:1985ki,Lee:1991ax}, these configurations arise in real fields and have no conserved charge, although see \cite{Kasuya:2002zs} for an adiabatic invariant. Their quantum mechanical decay rate was recently investigated in \cite{Hertzberg:2010yz}.
\footnote{They eventually decay either due to couplings to other field or their own outgoing radiation. The timescales depend on the model under consideration.} 

The presence of oscillons could have important consequences for the post inflationary universe. They could lead to an enhanced decay rate for the inflaton \cite{McDonald:2001iv} or explosive production of particles in localized regions \cite{McDonald:2001iv,Hertzberg:2010yz}, long after the homogenous inflaton has fragmented. Enhanced density perturbations due to oscillons provide seeds for formation of structure via accretion and mergers on much smaller scales than those responsible for current large scale structure. For large oscillons, there is a possibility of producing primodial black holes (for example, see \cite{Khlopov:2008qy}), which could in turn constrain the inflationary potential. In addition, it will be interesting to study gravitational wave production due to the formation, collapse and mergers of these lumps. A related investigation regarding gravitational waves from gravitationally collapsed inflaton clumps  was carried out recently in \cite{Jedamzik:2010hq} and from $Q-$balls in \cite{Kusenko:2008zm,Kusenko:2009cv, Chiba:2009zu,Enqvist:2003gh}. In models where the field fragments efficiently into oscillons, the gravitational wave spectrum from preheating \cite{Khlebnikov:1997di,Easther:2006gt,Easther:2006vd,Dufaux:2007pt,GarciaBellido:2007af,Easther:2007vj} could be modified. None of these consequences of oscillons have been properly explored in a cosmological setting.

An important first step towards exploring  the cosmological consequences of oscillons is estimating their number density, fraction of energy density in oscillons, and individual characteristics. In this paper, we undertake this task for a class of single field inflation models (or effectively single field models during the oscillatory phase). Near the minimum of the potential, the dependence on the inflaton $\varphi$, is assumed to be $V(\varphi)=m^2\varphi^2/2-\lambda\varphi^4/4+g^2m^{-2}\varphi^6/6 \hdots$. We require (i) $V'(\varphi)-m^2\varphi<0$ for some range of the field and (ii) $(\lambda/g)^2\ll1$.  Condition (i) is generic for any potential that flattens out at large field values, for example the potential in axion monodromy inflation \cite{Silverstein:2008sg} or models of hybrid inflation \cite{Linde:1993cn}. Note that (i) does not imply the need for an inflection point, whereas (ii) helps in protecting the oscillons from a collapse instability, though is not strictly required \cite{Amin:2010jq}. Throughout the analysis, we assume that the self interactions of the inflaton dominates over the coupling to other fields during the time of interest.  We discussed this model in detail in \cite{Amin:2010jq}, including an analysis of oscillon solutions in an expanding universe, and their stability.

We now provide a brief synopsis of our approach and the organization of the rest of the paper. Details about our choice of potential and associated assumptions are discussed in section \ref{sec:model}.  In section \ref{sec:icl} we start with zero point fluctuations in the inflaton during its oscillatory phase and follow their linear evolution analytically. Fluctuations in a limited band of wavenumbers get amplified via parametric resonance.  Non-linear effects that give rise to large oscillons only become important if the initial fluctuations are amplified sufficiently fast, Hubble expansion being the competing effect. We use this criterion to identify the values of $\lambda,g$ and $m$ for which we get copious production of oscillons. This linear analysis also allows us to determine an important scale, $k_\nl$, the scale that is the first to become nonlinear. We hypothesize that it is this scale that determines the co-moving number density of large oscillons:  $n_{osc}a^3\sim (k_{\nl}/2\pi)^3$. In section \ref{sec:nosc3d} we obtain $k_{\nl}$ and hence estimate the number density in terms of the parameters of the inflaton Lagrangian ($m,\lambda$ and $g$).

To test whether our estimate provides a good approximation for the number density, we need a lattice simulation of the field in an expanding universe. In this paper (section \ref{sec:1d}) we discuss a detailed set of numerical simulations in $1+1$-dimensions. We investigate the individual characteristics of oscillons, the fraction of energy density in oscillons, as well the number density of oscillons produced and compare them to our analytic estimates. We find good agreement between the two (better than a factor of 2) and the expected scaling as parameter values are changed. Results from numerical simulations in $3+1$-dimensions  will be presented in an upcoming publication \cite{Amin:2010}.
\footnote{Given the range of scales that need to be resolved ($H<k/a<m$ where $H\ll m$), the $3+1$ dimensional simulations are quite challenging.
Preliminary simulations in $3+1$-dimensions 
(run in collaboration with Richard Easther and Hal Finkel) are consistent with the ansatz presented in this paper to within a factor of $\sim 2$ \cite{Amin:2010}. In $3+1$-dimensions, oscillons also take up $>50\%$ of the energy density of the inflaton, as is the case in $1+1$-dimensions.}

Our conclusions and future directions are presented in section \ref{sec:disc}. The details of our numerical set-up, additional discussion of initial conditions and a derivation of the Floquet exponent are deferred to appendices. Animations based on our numerical simulations can be found online at \href{http://www.mit.edu/~mamin/oscillons.html}{http://www.mit.edu/$\sim$mamin/oscillons.html}. 

Before we begin our investigation, we briefly review the relevant literature on emergence of oscillons in the early universe.  This list is by no means exhaustive. Formation of ``axitons" in the axion field during the QCD phase transition was investigated in \cite{Kolb:1993hw}. Emergence of such pseudo-solitons (not all of them are oscillons) has been investigated in certain supersymmetric, hybrid inflation models \cite{McDonald:2001iv, Broadhead:2005hn, Copeland:2002ku}. The approach in \cite{McDonald:2001iv} is closest to the one taken in this paper. However, the potential considered in \cite{McDonald:2001iv} makes the oscillons susceptible to a collapse instability in $3+1$-dimensions (see \cite{Amin:2010jq}). Our method for calculating $k_\nl$ is somewhat different, specifically revealing the impact of the expansion rate on the number density. $Q$-ball production in the early universe shares many similarities with oscillon production \cite{Kusenko:2008zm,Kusenko:2009cv,Chiba:2009zu} , except, unlike oscillons, $Q$ balls are present in complex valued scalar field with an exactly conserved charge. Formation of oscillons from quasi-thermal initial conditions in a $1+1$ dimensional de-Sitter universe was investigated in \cite{Farhi:2007wj}. A numerical investigation of  oscillon-oscillon and oscillon-domain wall interactions in $2+1$ dimensions was carried out in \cite{Hindmarsh:2007jb}. Very recently, results of a $3+1$ dimensional numerical simulation of oscillons emerging from symmetry breaking phase transitions (with quasi-thermal initial conditions in deSitter space) was presented in \cite{Gleiser:2010qt}. They reported $\sim 2\%$ of the total energy density fraction in oscillons. To the best of our knowledge, an estimate for the number density in terms of the parameters of the model and its comparison with simulations was not provided in any of the above mentioned papers.  
 
\section{Our model and associated assumptions}
\label{sec:model}

We begin with the an action which includes the inflaton and  gravity ($\hbar=c=1$):
\begin{equation}
{S}=\int dx^4\sqrt{-\mathcal{G}} \left[\frac{\mpl^2}{2}R-\frac{1}{2}\partial^\mu\varphi\partial_\mu\varphi-V(\varphi)\right],
\end{equation}
where $\mathcal{G}$ is the determinant of the metric, $R$ is the Ricci scalar and $\mpl$ is the reduced Planck mass. We assume that
\begin{equation}
\label{eq:potential}
V(\varphi)=\frac{1}{2}m^2\varphi^2-\frac{\lambda}{4}\varphi^4+\frac{g^2}{6m^2}\varphi^6+\hdots
\end{equation} 
near the minimum of the potential (see figure \ref{fig:potential}). The effective mass of the inflaton at the bottom of the potential is $m$ whereas $\lambda$ and $g$ are dimensionless parameters.  
We will assume that $|\lambda|,g\ll1$ and $m/\mpl\ll1$. The fiducial values we have in mind are $m/\mpl\sim 5\times 10^{-6}$ and  $|\lambda|,g\sim 10^{-6}$. The typical field values of interest are $\varphi\ll m/\sqrt{\lambda}$. For such values, we assume that we can ignore terms  beyond $\varphi^6$. Furthermore, we assume that far away from the minimum, the potential is consistent with results from the cosmic microwave background anisotropies \cite{Komatsu:2008hk}. We note that although we assume $m/\mpl\approx 5\times 10^{-6}$ (chosen to be consistent with the amplitude of temperature fluctuations seen by WMAP \cite{Komatsu:2008hk}), its actual value is not necessarily set by the amplitude of the fluctuations. Inflation can take place for field values where the shape of the potential is unrelated to the shape at its minimum.
In addition, we make the following non-trivial assumptions about the potential near $\varphi=0$:

  \begin{figure}[tbp] 
     \centering
     \includegraphics[width=5in]{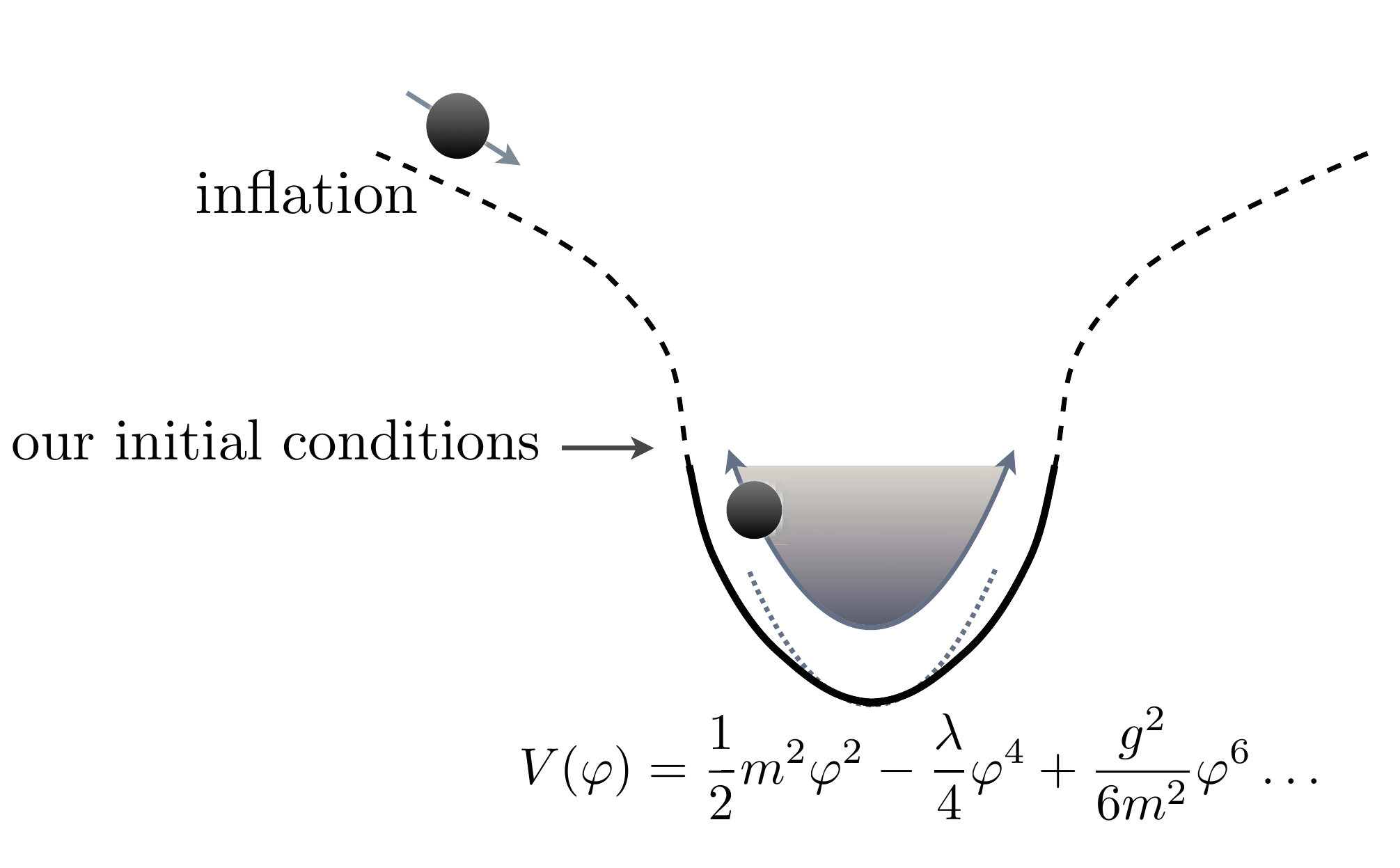} 
     \caption{The above figure shows an inflaton potential which can support oscillons. We are only interested in the shape near the bottom of the potential well. Near the minimum, the potential (thick black line) has to be shallower than quadratic (dotted line). The inflaton potential away from the minimum (dashed line)  is not crucial to our analysis apart from its possible influence on the initial conditions.}
     \label{fig:potential}
  \end{figure}


\begin{itemize}
\item $V'(\varphi)-m^2\varphi<0$ is required for some range of $\varphi$ for the existence of oscillons. This implies that $\lambda>0$ (ignoring terms beyond $\varphi^6$). Heuristically, the potential has to be shallower than quadratic near the minimum.
\item We assume that $(\lambda/g)^2\ll 1$. This is not strictly required, however it makes a semi-analytic analysis possible and allows for the existence of large ($R\gg m^{-1}$), massive ($M_{osc}\gg m$), robust flat-topped energy density configurations. In \cite{Amin:2010jq} we provided a detailed analysis of the oscillon solutions and their stability in the above class of models. The downside of this assumption is that it makes our model somewhat special. This ratio appears often throughout the paper. The reader can assume a value $\z^2\sim 10^{-1}$ while reading most of the text.
\item We assume that $\beta\equiv \sqrt{\lambda}\z(\mpl/m)\gg 1$. Heuristically, $\beta\sim \mu/H$ characterizes the growth rate of fluctuations $\mu$ compared to the Hubble rate $H$. This condition on  $\beta$ is necessary for the parametric amplification of the initial fluctuations that ultimately form large oscillons. We will discuss this requirement after an analysis of the linearized solutions. While reading through the paper, the reader can assume $\beta\sim 10^2$. We keep $(m/\mpl)$ fixed in this paper, hence $\beta\sim10^2$ and $\z^2\sim10^{-1}$ is equivalent to $\lambda \sim 2.5\times 10^{-6}$. When we vary parameters, we find it convenient to treat  $\z$ and $\lambda$ as independent parameters instead of $\lambda$ and $g$.

\item The coupling of the inflaton to the other fields (including the Standard Model ones) is assumed to be small compared to its self-interactions. This does not mean that the inflaton does not decay, but that significant decay happens long after the fragmentation of the inflaton. Since we only require inflaton decay and thermalization by the time BBN begins, this is easily achievable. Without this assumption, or specifying the nature of coupling to other fields, we cannot make any concrete statements about what fraction of the inflaton energy density ends up in oscillons.
 \end{itemize}

We assume that $V(0)=0$. For concreteness, from now on we will discuss the emergence of oscillons in the context of \eqref{eq:potential}, with terms beyond $\varphi^6$ being explicitly set to zero. However, we stress that our techniques are general and can easily be applied to a wider class of models.

  \section{Initial conditions and linear evolution}
    \label{sec:icl}

In this section we follow the evolution of linear fluctuations as they get amplified via parametric resonance. For a linear analysis of the fluctuations, it is easy to obtain the solutions numerically. However, we choose to provide a discussion based on approximate, analytic solutions to reveal the effects of different parameters on our results. 

The equation of motion for the inflaton is
 \begin{equation}
 \begin{aligned}
\Box \varphi=V'(\varphi).
 \end{aligned}
 \end{equation} 
 We write $\varphi(t,{\bf{x}})$ as a sum of the homogeneous piece $\bp(t)$ and fluctuations $\delta \varphi(t,{\bf{x}})$. We assume that $\langle\delta\varphi(t,{\bf{x}})\rangle=0$ where $\langle\hdots\rangle$ denotes spatial averaging. Similarly the metric is FRW with small deviations : $\mathcal{G}_{\mu\nu}dx^\mu dx^\nu=\left(\mathcal{G}^{B}_{\mu\nu}+\delta \mathcal{G}^{\mu\nu}\right)dx^\mu dx^\nu$ where $\mathcal{G}^{B}_{\mu\nu}dx^\mu dx^\nu=-dt^2+a^2(t)d{\bf{x}}\cdot d{\bf{x}}$ and $\delta \mathcal{G}_{\mu\nu}(t,{\bf{x}})\ll \mathcal{G}^{B}_{\mu\nu}(t)$. For the following, $t$ is cosmic time in an FRW universe and ${\bf{x}}$ is a co-moving, cartesian co-ordinate on a fixed time slice.
 
 \subsection{Homogenous background evolution}
The equations of motion for the homogeneous field are
\begin{equation}
 \begin{aligned}
& \partial_{t}^2\bp+3H \partial_{t}\bp+V'(\bp)=0,\\
& H^2=\frac{1}{3\mpl^2}\left[ \frac{1}{2}\partial_{t}^2\bp+V(\bp)\right],
 \end{aligned}
 \end{equation}
 where $H=\dot{a}(t)/a(t)$. We have assumed that the backreaction of the fluctuations on the homogeneous equations of motion is small. That is, $V'''(\bp)\langle\delta\varphi^2\rangle\ll V'(\bp)$. 
 After the end of inflation, for the potential in \eqref{eq:potential}, the field oscillates about the minimum with a decaying amplitude. At the bottom of the potential, the solution is well approximated by
 \begin{equation}
 \begin{aligned}
&\bp(t)\approx \frac{\bp_i}{\sqrt{a^3(t)}}\cos(\omega t),\\
& H\approx\frac{H_i}{\sqrt{a^3(t)}},
 \end{aligned}
 \label{eq:homsol}
 \end{equation}
 where frequency of oscillation is given by $\omega^2\approx(m^2-\frac{3\lambda}{4}\bp^2+\hdots)$. This solution assumes that $\bp$ is small enough for the non-linear terms in the potential  to be sub-dominant. We have defined $a(t_i)=a_i=1, \bp(t_i)=\bp_i$ and $H(t_i)=H_i$.  It is convenient to chose $\bp_i$ and $H_i$ (and hence $t_i$) based on the structure of the instability bands for the fluctuations. The time $t_i$ is chosen so that before $t_i$, there is no significant field induced amplification of fluctuations on scales of interest for oscillon formation. We will discuss this in a later part of this section [see equation \eqref{eq:initialfield}]. For the moment, we assume that $H_i\ll m$ and $m\ll\bp_i\ll \mpl$.

 \subsection{Linear evolution of fluctuations}
 
\noindent The linearized equation of motion for the field fluctuations are:
 \begin{equation}
 \begin{aligned}
 \partial_{t}^2\delta\varphi+3H \partial_{t}\delta\varphi+\left(-\frac{\nabla^2}{a^2(t)}+V''(\bp)\right)\delta\varphi=F\left[\delta\mathcal{G}_{\mu\nu}\right],
  \end{aligned}
 \label{eq:LinearPosition}
 \end{equation}
where r.h.s arises due to fluctuations of the metric. Note that the linearized expression on the l.h.s is justified as long as  $V''(\bp)\gg V'''(\bp)\delta\varphi$\footnote{We thank Raphael Flauger for pointing this out}.
In Fourier space we have
 \begin{equation}
 \begin{aligned}
  \label{eq:LinearFourier}
 \partial_{t}^2\delta\varphi_{k}+3H \partial_{t}\delta\varphi_{k}+\left(\frac{k^2}{a^2(t)}+V''(\bp)\right)\delta\varphi_k
 =F_k\left[\delta\mathcal{G}_{\mu\nu}\right].
 \end{aligned}
 \end{equation}
The solution to the above equation depends strongly on the relationship between $k, H$ and $V''(\bp)$. When the homogeneous field is in the oscillatory regime, $V''(\bp)-m^2$ and $F\left[\delta\mathcal{G}_{\mu\nu}\right]$ are oscillatory and could give rise to parametric amplification of the fluctuations. Growth of $\delta\varphi$ was also investigated in \cite{Johnson:2008se} for a similar system  in the late universe (without the $\varphi^6$ term).

Let us first consider the case where $V''(\bp)-m^2\ll F\left[\delta\mathcal{G}_{\mu\nu}\right]$. As a concrete example, consider $V(\varphi)=m^2\varphi^2/2$. This case has been recently investigated in detail in \cite{Jedamzik:2010dq} and \cite{Easther:2010mr}. On the other hand if $V''(\bp)-m^2\gg F\left[\delta\mathcal{G}_{\mu\nu}\right]$ we can ignore the gravitational effects and concentrate on the nature of the field induced resonance \footnote{This is a scale and time dependent statement, and one has to look at the structure of the resonance bands to confirm this. We expect the metric fluctuations to play a role in setting up the initial conditions and also after the field resonance stops being efficient on scales $k/a\lesssim \sqrt{3Hm}$ \cite{Jedamzik:2010dq,Easther:2010mr}.}. We wish to concentrate on this regime. 

\subsubsection{Initial Conditions for fluctuations}

We are interested in sub-horizon scales $k/a\gg H$. The initial conditions at $t=t_i$ depend on the potential $V(\bp)$ beyond $\bp(t_i)$ and long wavelength ($k\ll m$) gravitational effects. For example, the amplitude of modes that never left the horizon compared to ones that did so during inflation can be significantly different at $t_i$. Even if we assume that field induced resonance only becomes important after $t_i$ [this is true for our potential in \eqref{eq:potential}], without specifying the detailed shape of the inflaton potential beyond $\bp_i$, it is difficult to estimate these effects. However, all of these effects are likely to increase the level of fluctuations  compared to the zero point fluctuations at any given sub-horizon scale. We take a somewhat conservative approach in choosing our initial conditions, taking them to be consistent with the zero point flucutations of the field in Minkowski space. Treating these fluctuations as a classical, Gaussian, random field, the typical value of fluctuation on a scale $k^{-1}$ is given by (ignoring interaction terms):
\begin{equation}
k^{3/2}\delta\varphi_k\approx \frac{k^{3/2}}{\sqrt{2\omega_k}}
\end{equation}
where $\omega_k^2=k^2+m^2+\mathcal{O}[H_i^2]$ (see for example \cite{QFTBook:2007}). Since we assume that $H_i\ll m$,  we will ignore the $H_i$ piece. As we will see, parametric resonance is strong enough to generate oscillons from these conservative initial conditions. We are treating these initial sub-horizon fluctuations classically, since we expect their occupation numbers to be large after they hit the instability band \cite{Khlebnikov:1996mc}. See appendix A for further details on the initial conditions and some justifications. 
\subsubsection{Resonance in Minkowski space}
\begin{figure}[tbp] 
     \centering
     \includegraphics[width=5.5in]{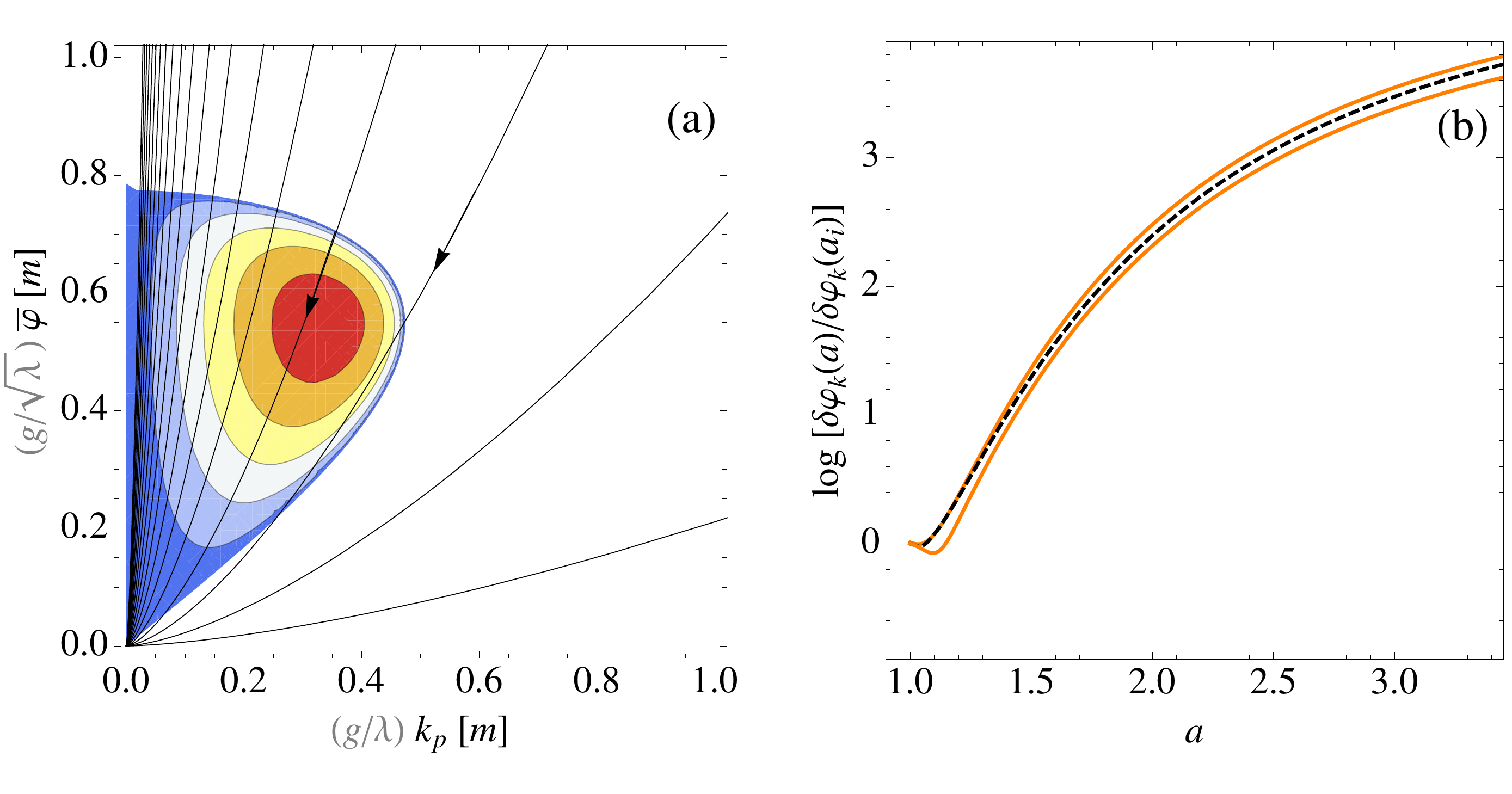} 
     \caption{In figure (a), the resonance band for the potential $V(\bp)=m^2\bp^2/2-\lambda\bp^4/4+g^2\bp^6/6$ is shown. The shaded region has a positive Floquet exponent. The thin black-lines show the ``path" following by modes as the universe expands. The amplification of fluctuations can be estimated by integrating the Floquet exponent along these ``paths". Our initial conditions are chosen at the dashed line. In figure (b), the amplification of a mode with $(g/\lambda)k=0.3m$ is shown as a function of the scalefactor $a$. The dashed line is our analytic estimate. The orange lines are the two numerically integrated solutions with orthogonal initial conditions: $\delta\varphi_k=0,1$ and $\partial_t\delta\varphi_k=1,0$. Also see discussion following equation \eqref{eq:evolfluc}. The waveneumber and field are expressed in units of the mass $m$.}
     \label{fig:floquet}
  \end{figure}

The oscillatory terms of $V''(\bp)-m^2$ in equation \eqref{eq:LinearFourier} can lead to parametric amplification of fluctuations for some selected bands of wavenumbers and values of the background field $\bp$. It is these resonantly amplified modes which form oscillons.  In an instability band, if we were to ignore expansion ($a=1,H=0$) we can treat the evolution of the fluctuations via standard Floquet analysis (for example, see \cite{Hill:1979}). The most unstable modes have a solution of the form:
\begin{equation}
\delta\varphi_k(t)\propto e^{\mu_k t}P(t),
\end{equation}
where $\mu_k$ is the Floquet exponent and $P$ is a periodic function. The Floquet exponent depends on the amplitude of the ``pump" field $\bp$ as well as the wavenumber $k$. In figure \ref{fig:floquet}(a), we show the Floquet exponent as a  function of the amplitude $\bp$ and wavenumber. The colored regions represent the first instability band, with the color representing the real part of the Floquet exponent (red corresponds to a large Floquet exponent). Note that this is different from the usual Mathieu instability chart, where the Floquet exponents are plotted in terms of the resonance parameter $q\propto \bp^2$ and $A_k=(k/m)^2+2q$. Higher order, narrow bands [$\Delta k\lesssim (\lambda/g)^2)$] exist at $k>m$, for example at $k\sim\sqrt{3}m$, beyond the right edge of the plot.

Under the assumption that $(\lambda/g)^2 \ll 1$ we can write down an approximate form of the Floquet exponent shown in figure \ref{fig:floquet}(a) as 
\begin{equation}
\begin{aligned}
\label{eq:nxfloquet}
&\mu_{k}(\bar{\varphi})\approx \frac{k}{2}\sqrt{\frac{3\lambda}{2}\left(\frac{\bar{\varphi}}{m}\right)^{\!2}\left[1-\left(\frac{\bp}{\bp_i}\right)^2\right]-\left(\frac{k}{m}\right)^{\!2}},\\
\end{aligned}
\end{equation}
where 
\begin{equation}
\label{eq:initialfield}
\bp_i=\sqrt{\frac{3\lambda}{5g^2}} m
\end{equation}
is the amplitude where this band shuts off. The derivation can be found in appendix B .  For the parameters of interest, we have checked that this expression agrees with the numerically calculated Floquet exponent to  $\lesssim 1$ percent, except in a very tiny sliver at the edges of the Floquet band.


\subsubsection{Resonance in an expanding background}
To understand parametric resonance in an expanding background, we make the following identifications: 
\begin{equation}
\begin{aligned}
\label{eq:id}
&k\rightarrow k_\p=ka^{-1}(t),\\
&\bp\rightarrow\bp_ia^{-3/2}(t).
\end{aligned}
\end{equation}
 Here $k_\p$ is the physical wavenumber and $\bp_i$ was defined in equation \eqref{eq:initialfield}.  
 For future convenience, note that at $\bp=\bp_i$, $H_i\approx\sqrt{{\lambda}/{10g^2}}\left({m}/{\mpl}\right)m$.
 
 The identification \eqref{eq:id} defines a trajectory in the $k_\p-\bp$ plane [thin black lines in figure \ref{fig:floquet}(a)]. The expression for $\mu_k(a)$ along such a trajectory is 
\begin{equation}
\begin{aligned}
\mu_k(a)\approx \frac{1}{2}\frac{k}{a}\sqrt{\frac{9}{10}\frac{\z^2}{a^3}\left(1-\frac{1}{a^3}\right)-\left(\frac{k}{am}\right)^2}.
\end{aligned}
\end{equation}
A mode passing through the instability band can get amplified. Such modes are bounded by:
\begin{equation}
\frac{H_i}{\sqrt{a}}\ll k\lesssim 0.65\,\left(\frac{\lambda}{g}\right) m.
\end{equation}
The lower bound comes from insisting that the modes must be sub-horizon (which is a time dependent statement), whereas the upper bound comes from the requirement that they pass through the instability band [see figure \ref{fig:floquet}(a)]. 

The approximate amount of amplification undergone by a given mode is obtained by integrating $\mu_k$ along the corresponding trajectory in the $k_\p-\bp$ plane. 
Hence an approximate expression for the evolution of the amplified modes modes is 

\begin{equation}
\label{eq:evolfluc}
\delta\varphi_k(t)\sim \frac{\delta\varphi_k(t_i)}{a^{3/2}(t)}\exp\left[\int_{\partial(t,k)} d\tau \mu_k(\tau)\right]
=\frac{1}{a^{3/2}}\frac{1}{\sqrt{2\omega_k}}\exp\left[\int_{\partial(a,k)} d\ln \bar{a} \frac{\mu_k(\bar{a})}{H(\bar{a})} \right].
\end{equation}

We use the scalefactor as a time co-ordinate. The boundary of the integral $\partial(k,a)$ is obtained from $\mu(k,a)=0$. We are effectively assuming that the time scale of oscillation ($\sim m^{-1}$) is much shorter than $H^{-1}$. This expression should be used with caution. Its accuracy depends on the $k$ mode under consideration and should always be checked with a direct numerical integration (also see \cite{Greene:1998pb}). In figure \ref{fig:floquet}(b), we compare the above expression with the numerical results for a particular $k$ mode: $k= 0.3(\lambda/g)m$, with $(\lambda/g)^2=0.2$ and $\lambda=3.13\times 10^{-7}$.  \footnote{Ideally we would compare the numerically calculated eigenvector of the time dependent Floquet matrix with our approximate expression \eqref{eq:evolfluc}. However, this would require diagonalizing the Floquet matrix at every time step. Here the comparison is made between our expression \eqref{eq:evolfluc} and the fundamental solutions. For the concerned reader, we note that our main result, regarding the number density, will depend on $\log k^{3/2}\delta\varphi_k$ and not on $k^{3/2}\delta\varphi_k$.} 
\subsubsection{Condition of significant amplification}
From the exponent in equation \eqref{eq:evolfluc} it is clear that we need $\mu_k(a)\gg H$ for a significant amplification of the fluctuations.  This means that:
\begin{equation}
\frac{\mu_k(a)}{H}=\frac{\mpl}{m}\sqrt{\lambda}\left(\frac{\lambda}{g}\right)\left[\sqrt{\frac{5}{2}}\sqrt{\frac{9}{10}\frac{\tk^2}{a^2}\left(1-\frac{1}{a^3}\right)-\left(\frac{\tk^4}{a}\right)}\right]\gg1.
\label{eq:beta}
\end{equation}
where $\tk\equiv (g/\lambda m)k$. This motivated our definition $\beta=\lambda^{3/2}/g(\mpl/m)$ in section \ref{sec:model}. For appropriate values of $\lambda,g$ and $m$, we can get $\beta$ and hence $\mu_k/H\gg1$. However, note that for any $\beta$ and $\tk$, the amplification will cease eventually as $a\gg1$. The amplitude can also stop increasing if the mode leaves the instability band, or when non-linearities cap off the growth. In addition, the gravitational terms from the r.h.s of equation \eqref{eq:LinearPosition} might become important.

Putting all these results together, the evolution of such fluctuations after they hit the instability band is
\begin{equation}
\label{eq:fullsol}
\delta\varphi_k(a)\sim \frac{1}{\sqrt{2\omega_k}}\frac{1}{a^{3/2}}\exp\left[ \beta f(\tilde{k},a)\right],
\end{equation}
where
\begin{equation}\
f(\tilde{k},a)=\sqrt{\frac{5}{2}}\int_{\frac{a^3-1}{a}\ge\frac{10}{9}\tilde{k}^2} d\ln \bar{a}\left[\tilde{k}\sqrt{\frac{9}{10\bar{a}^2}\left(1-\frac{1}{\bar{a}^3}\right)-\frac{\tilde{k}^2}{\bar{a}}}\right].
\end{equation}
This integral can easily be done numerically and for $a\sim \textrm{a few}$, the integral peaks at $\tk\sim 0.4$. The typical modes that are amplified have a co-moving wavenumber $k\lesssim m\z$, with a corresponding Floquet exponent $\mu\sim 0.1m\z^2$. Recall that we assumed $\z^2\sim 0.1$ and we require $\beta=\sqrt{\lambda}\z(\mpl/m)\gtrsim 10^2$ for significant amplification. The Hubble parameter at the time of amplification is $H\sim m\z^2/\beta\sim 10^{-3}m$. Although these simple estimates serve as a useful guide, we use the solution presented in equation \eqref{eq:fullsol} for the next section.

Before we end this section we would like to visit the assumptions we made during our analysis. First, we have concentrated on the first instability band and have ignored the higher order bands (for example at $k_\p\sim\sqrt{3}m$). The higher order bands are very narrow  ($\Delta k_\p \lesssim(\lambda/g)^2$) and, due to expansion, modes rapidly redshift through them. The structure of these higher order bands is rather non-trivial. Their width in the $k_\p-\bp$ plane remains small, though it does increase with the amplitude before shutting off. Oscillons have sizes significantly larger than $m^{-1}$. As a result, we do not expect these high $k_\p$ modes to significantly affect oscillon formation. Nevertheless, if modes pass through a large number of such bands or spend a long time in one of these narrow bands (that is, $H$ is very small) large amplification is possible. These fluctuations could lead to oscillon formation if their wavelengths redshift by a sufficient amount. This would significantly complicate the analysis, and is not taken into account here. 

Second, we ignored terms beyond $\varphi^6$ in the potential. This can change the structure of the resonance bands (especially for $\bp>\bp_i$). However, one can always obtain the resonance bands numerically by including the extra terms and repeating the above analysis. 

Our linear analysis of the growth of fluctuations eventually breaks down as the amplitude of the fluctuations $\delta\varphi\sim k^{3/2}\delta\varphi_k$, becomes comparable $\bp$. At this point, we no longer have a homogeneous pump field to parametrically amplify the fluctuations. In addition, as $\delta\varphi$ approaches $V''(\varphi)/V'''(\varphi)$, we have to take into account the interaction of different $k$ modes. The modes start interacting rapidly with each other, ultimately forming oscillons. This process is difficult to follow analytically, however as we will see, the number of oscillons can be predicted based on the linear analysis. 

 
   \section{Estimating the number density of oscillons}
   \label{sec:nosc3d}
    In this section, we estimate the number of oscillons that emerge from the breakup of the inflaton at the end of inflation. The idea is to use our understanding of the linear evolution to predict the number density of oscillons. 

Oscillons emerge from the parametrically amplified fluctuations. As parametric resonance ceases to be efficient, there is a characteristic length scale  where the fluctuations in the field are highly nonlinear. We claim that the number density of oscillons can be estimated using:
 \begin{equation}
 n_{\textrm{osc}}a^3\sim \left(\frac{k_{\nl}}{2\pi}\right)^3
 \end{equation}
where $k_{\nl}$ label the modes that become non-linear first. More explicitly, these are the modes for which the condition,
\begin{equation}
\label{eq:nlcon}
k^{3/2}\delta\varphi_k\sim \bp 
\end{equation} 
is satisfied the earliest. We admit, that this is somewhat ad-hoc. One could have also used $\delta\varphi\sim V''(\bp)/V'''(\bp)$ (mode-coupling) or $\delta\varphi\sim \sqrt{V'(\bp)/V'''(\bp)}$ (backreaction on homogeneous evolution) as a condition for obtaining a slightly different $k_\nl$. We use equation \eqref{eq:nlcon} because for the model under consideration, $\delta\varphi\sim\bp$ happens before $\delta\varphi\sim V''(\bp)/V'''(\bp)$ or $\delta\varphi\sim \sqrt{V'(\bp)/V'''(\bp)}$ are satisfied. Nevertheless, we have checked that using the other conditions yields similar results for our estimate of the number density\footnote{A more accurate analysis would require answering a rather difficult question:  Why are oscillons attractors in the space of many possible field configurations?}. Heuristically, we are merely counting the number of large peaks in the energy density. The linear analysis tells us how many such peaks we should expect.
 
Using $k^{3/2}\delta\varphi_k\sim (2\omega_k)^{-1/2}(k/a)^{3/2}\exp[\beta f(\tk,a)]$ and $\bp=\bp_i a^{-3/2}$ in \eqref{eq:nlcon} we get:
\begin{equation}
\label{anlcon3d}
\frac{\tk^{3/2}}{\sqrt{2\omega_{\tk}}} \exp\left[ \beta f(\tk,a_{\nl})\right]=\sqrt{\frac{3g}{5\lambda^2}}
\end{equation}
where $\omega_{\tk}=\sqrt{1+\z^2\tk^2}$, $\tk=m^{-1}\z k$ and $\beta=\sqrt{\lambda}\z(\mpl/m)$. Different modes will become nonlinear at different times or scalefactors, $a_{\nl}(\tk)$. We can obtain the modes that become non-linear first by solving for $\tk$ in
\begin{equation}
\label{knlcon3d}
\partial_{\tk} a_{\nl}(\tk_{\nl})=0.
\end{equation}

\begin{figure}[t] 
     \centering
     \includegraphics[width=5.5in]{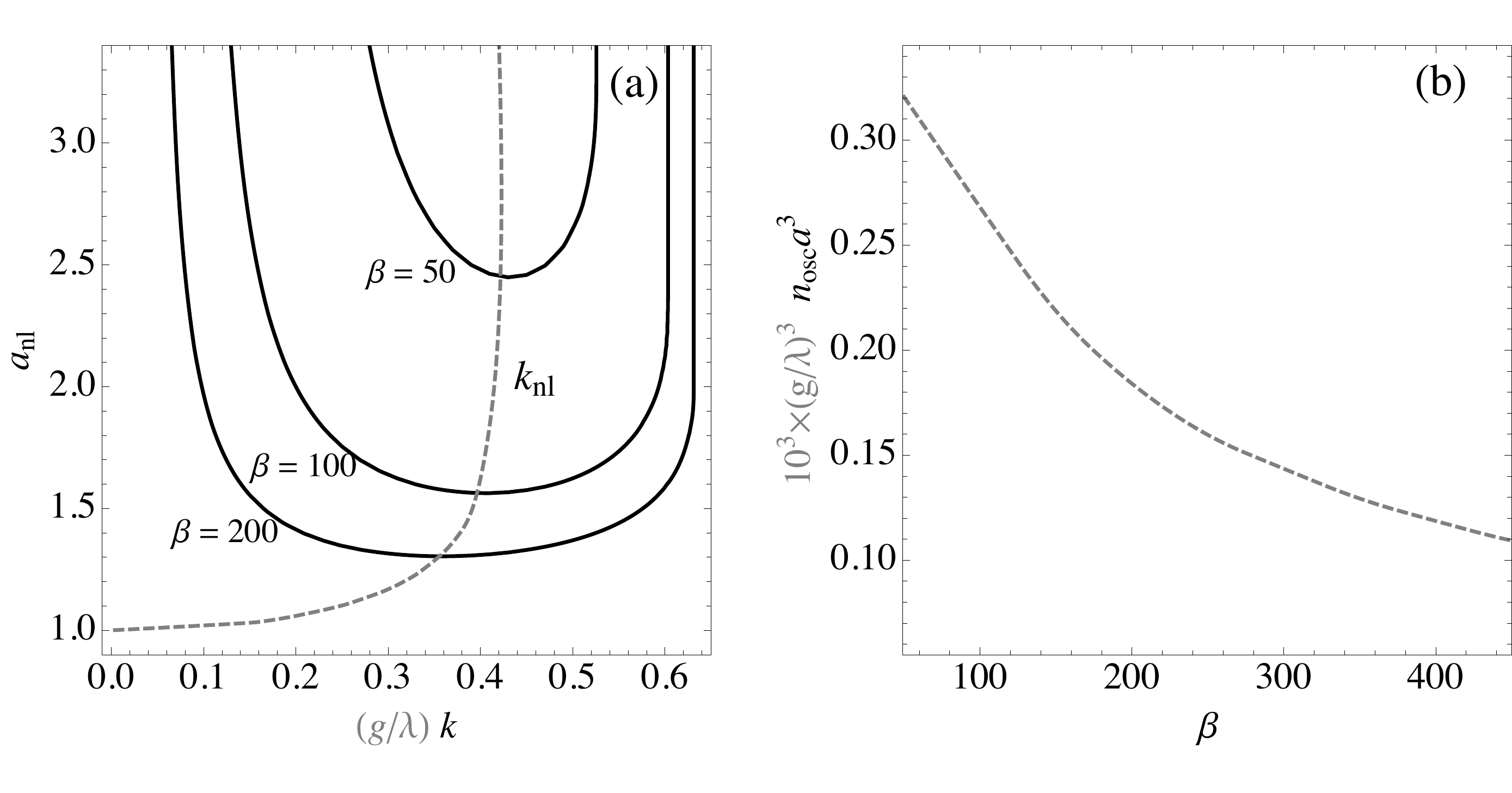} 
     \caption{[$3+1$-dimensions] In (a), the solid curves show the scalefactor at which a mode with given $k$ becomes non-linear. Different curves correspond to different values of $\beta=\sqrt{\lambda}\z(\mpl/m)\sim \mu/H$ where $\mu$ is the Floquet exponent and $H$ is the Hubble parameter. The dashed curve in $(a)$ represents the modes that are the first to become non-linear. For the plot we chose $m/\mpl=5\times10^{-6}$ and $\z^2=0.2$ which implies $\lambda=1.25\times 10^{-10}\beta^2$. Based on figure (a),  in figure (b) we plot  the co-moving number density of large oscillons as a function of $\beta$ (or $\lambda$). As expansion gets slower, the number density of large oscillons should decrease. Spacetime variables are expressed in units of inverse mass $m^{-1}$.}
     \label{fig:nlfirst3d}
  \end{figure}

A plot of $a_{\nl}$ vs. $\tk$ is shown in figure \ref{fig:nlfirst3d}(a) for different values of $\beta$ with $\z^2=0.2$ and $m/\mpl=5\times 10^{-6}$. The dashed line corresponds to $\tk_{\nl}$, these are modes that become non-linear first. Since here we have fixed $m/\mpl$ and $\lambda/g$, we can treat $\beta$ or $\lambda$ as an independent variable. They are related by $\lambda=1.25\times 10^{-10}\beta^2$. Note that we could have equally considered $m/\mpl$ as the independent variable with fixed $\lambda$ and $g$. We prefer $\beta$ because it  controls how quickly modes get amplified. 

From figure \ref{fig:nlfirst3d}(a) we see that for a fixed $\z$, as $\beta$ increases, $k_{\nl}$ becomes smaller. This implies that the comoving number density of oscillons decreases with increasing $\beta$.  Although equations (\ref{anlcon3d}) and (\ref{knlcon3d}) are difficult  to solve analytically, one can easily estimate $k_{\nl}$ using figure \ref{fig:nlfirst3d}(a). For example for $\beta=10^2$ (or $\lambda=2.5\times 10^{-6}$) we get $k_{\nl}\approx 0.4m \z$.  This yields a co-moving number density of 
\begin{equation}
n_{\textrm{osc}}a^3\sim \left(\frac{0.4}{2\pi}\frac{\lambda}{g} m\right)^3.
\end{equation}
Note that as $\beta$ gets large, the $a_{\nl}$ vs. $\tk$ curve becomes exceedingly flat near its minimum. As a result, $k_{\nl}$ is not sharply defined. In this regime, our ansatz is like to fail. 

Let us understand the effects of different parameters on the estimate for the number density. To a good approximation, $k_{\nl}$ obtained by solving equations \eqref{anlcon3d} and \eqref{knlcon3d} is given by:
\begin{equation}
k_{\nl}\sim \beta^{-1/5}\z m.
\end{equation}
Thus, the number density of oscillons is given by
\begin{equation}
 n_{\textrm{osc}}a^3\sim \beta^{-3/5}\left(\frac{\lambda}{g}\frac{m}{2\pi}\right)^3,
 \label{eq:nosc3d}
 \end{equation}
where $\beta=\sqrt{\lambda}\z(\mpl/m)$. This is the main result of this section. We have checked that this is consistent ($\sim15\%$) with a numerical solution of equations \eqref{anlcon3d} and \eqref{knlcon3d} in the range  $\z^{-2}[5-10]$ and $\lambda[6.25\times 10^{-7}-6.25\times 10^{-5}]$. We also varied $m/\mpl$ within an order of magnitude of $5\times 10^{-6}$ and found similar results.\footnote{Better fits can be found, but this is good enough for our purposes.} 
 
 One might wonder if it was possible to read off $k_\nl$  from figure \ref{fig:floquet}(a) directly. The scaling with $\z$ comes from $k$, however the dependence on $\beta$ is somewhat difficult to see. Recall that $\beta$ characterized the growth rate of fluctuations $\mu$ compared to the Hubble rate $H$. As $\beta$ gets larger (equivalently $H$ gets smaller with $\z$ fixed), the low momentum modes get amplified before the relatively higher $k$ modes become resonant (see shape of instability band in figure \ref{fig:floquet}). As a result, the non-linearity condition \eqref{eq:nlcon} is satisfied by longer wavelength modes first.

Although plausible, equation \eqref{eq:nosc3d} is an approximation and should be checked with detailed numerical simulations. We have not discussed some important aspects related to the emergence of oscillons. Non-linear interactions between oscillons is not fully understood. Their evolution following their emergence, as they merge and scatter is difficult to tract analytically. In particular, for a fixed $\z$, as the Hubble expansion gets smaller, ($\beta\gg 100$, equivavently $\lambda\gg 10^{-6}$ or we increase $\mpl/m$ significantly), oscillon-oscillon interaction as well as interactions between oscillons and large non-linearities can alter the number density. 

\section{Emergence of oscillons: $1+1$-dimensions}
\label{sec:1d}
In the previous sections we provided an ansatz for the number density of oscillons produced at the end of inflation in $3+1$-dimensions. Ideally, we would like to test this ansatz through numerical evolution of the fully non-linear system on a lattice and test explore a wide range of parameter space. However, a large dynamic range of scales is necessary to resolve a significant fraction of the Hubble horizon as well as the internal structure of oscillons in an expanding universe. These simulations are time and memory intensive, and will be presented in an upcoming publication \cite{Amin:2010}.

In this paper, we test our analysis using $1+1$-dimensional numerical simulations. This significantly simplifies the numerics as well as the analytical expressions. 
We first reduce the $3+1$-dimensional analysis to $1+1$-dimensions and then proceed towards the comparison with numerical simulations. 
\subsection{Linear evolution and initial conditions: $1+1$-dimensions}
The potential for the inflaton field is taken to be 
\begin{equation}
V(\varphi)=m^2\left[\frac{1}{2}\varphi^2-\frac{\lambda}{4}\varphi^4+\frac{g^2}{6}\varphi^6\right].
\end{equation}
Here $\hbar=c=1$. Note that in comparison with equation \eqref{eq:potential}, we have scaled out $m$ because in $1+1$-dimensions the field $\varphi$ is dimensionless. The equations of motion in a $1+1$-dimensional, homogeneous expanding universe are
\begin{equation}
\partial_t^2\varphi+H\partial_t\varphi-a^{-2}\partial_x^2\varphi+V'(\varphi)=0,
\end{equation}
where $H=\dot{a}(t)/a(t)$. In $1+1$-dimensions, in the oscillatory phase, the homogeneous field evolves as $\bp(a)=\bp_ia^{-1/2}$ and we assume that $H\approx H_i a^{-1/2}$. The choice of $H$ amounts to a prescription for the expansion history since in $1+1$-dimensions the Einstein tensor is identically zero. The structure of the Floquet instability band in terms of $\bp$ and $k_p$ does not change. Hence we take $\bp_i=\sqrt{3\lambda/5g^2}$ 
in analogy with the $3+1$ dimensional case. Although the structure in $\bp-k_p$ plane is unchanged,  the ``path" traced by a fluctuation with a given wavenumber does change due to the different scaling of $\bp$ and $H$ with $a$.   

    \begin{figure}[t] 
     \centering
       \includegraphics[width=5.5in]{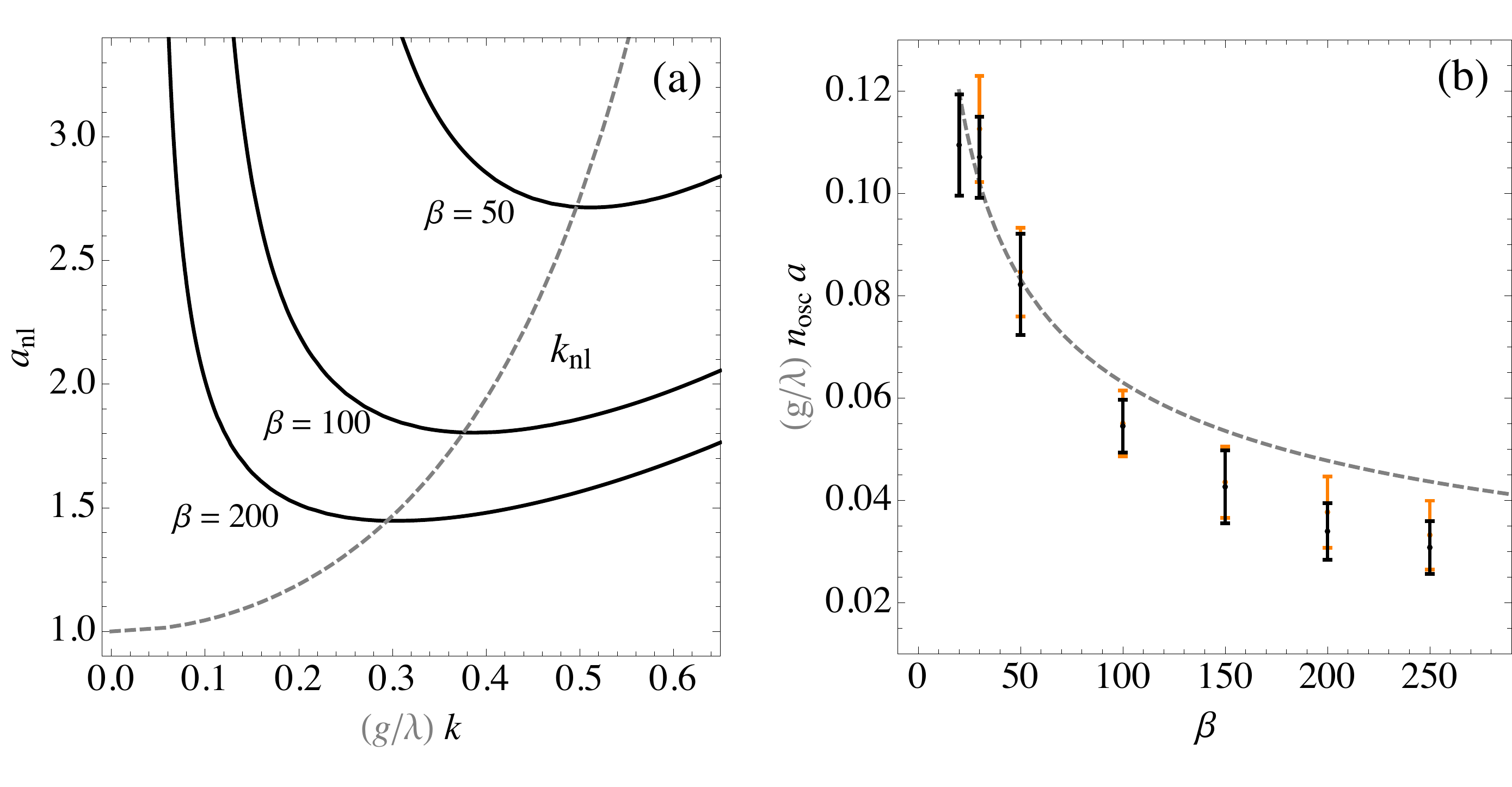} 
     \caption{[$1+1$-dimensions] In (a), the solid curves show the scalefactor at which a mode with given $k$ becomes non-linear. Different curves correspond to different values of $\beta=\sqrt{\lambda}\z(\mpl/m)\sim \mu/H$ where $\mu$ is the Floquet exponent and $H$ is the Hubble parameter. The dashed curve in $(a)$ represents the modes that are the first to become non-linear. For the plot we chose $m/\mpl=5\times10^{-6}$ and $\z^2=0.2$ which implies $\lambda\sim 10^{-10}\beta^2$. In figure (b) we plot our estimate for  the co-moving number density of large oscillons as a function of $\beta$ (dashed line). The orange [$(\lambda/g)^2=0.1$] and black $[(\lambda/g)^2=0.2]$ data points and 2 sigma error bars are from our $1+1$ dimensional numerical simulations, each point based on 10 independent realizations of the initial conditions.  The dependence on $\lambda/g$ is also captured by our estimate. The number density decreases with $H$. Note that $\lambda (m/\mpl)$ is being varied over $2(1)$ orders of magnitude. Spacetime variables are expressed in unit of the inverse mass $m^{-1}$.}
     \label{fig:nlfirst1d}
  \end{figure}

For $a\ge1$, the perturbations in the field evolve as [compare with equation \eqref{eq:fullsol}]:
\begin{equation}
\label{fSol1d}
k^{1/2}\delta\varphi_k(a)\sim \sqrt{\frac{\z}{2\omega_{\tk}}}\left(\frac{\tk}{a}\right)^{1/2}\exp\left[ \beta\frac{\tk}{a_{\tk}}\left(1-\frac{a_{\tk}}{a}\right)^{3/2}\right],
\end{equation}
where  $\omega_{\tk}=\sqrt{1+\zeta^2\tk^2}$, $\tilde{k}=m^{-1}(g/\lambda)k$ and $a_{\tk}=1+(10/9)\tk^2$. We have assumed  that $H_i=\sqrt{\lambda/10g^2}(m/\mpl)m=m\sqrt{1/10}\z^2/\beta$. We have chosen initial conditions at $a=1$ to be identical to the $3+1$ dimensional case. In contrast to the $3+1$ dimensional case, we were able to integrate $\int \mu_k(t)dt$ analytically.
\subsection{Number density: 1+1 dimensions}
As discussed before, oscillons form from the parametrically amplified modes. The modes that become non-linear earliest are the ones that satisfy the condition $\delta\varphi\gtrsim \bp$ first.  The scalefactor $a_\nl$ at which a given mode becomes non-linear and the co-moving wavenumber of the mode that becomes non-linear, $k_{\nl}$, can be obtained from
\begin{equation}
\label{nlcon1d}
\begin{aligned}
&a_{\nl}(\tk)= \frac{a_{\tk}}{1-\left(\frac{a_{\tk}}{\tk\beta}\ln\left[\sqrt{\frac{6\omega_{\tk}}{5\tk \lambda}\frac{\lambda}{g}}\right]\right)^{2/3}},\\
&\partial_{\tk}a_{\nl}(\tk_{\nl})=0.
\end{aligned}
\end{equation}
Repeating the procedure outlined in the $3+1$ dimensional case, in figure \ref{fig:nlfirst1d} we show $a_{\nl}(\tk)$ and $\tk_{\nl}$ for different values of $\beta\sim 10^{10}\sqrt{\lambda}$. We have fixed $\z^2=0.2$ and $m/\mpl=5\times10^{-6}$, however, the variation in $\z$ is essentially captured by our scaling of the horizontal axes. Unlike the $3+1$ dimensional case, $a_\nl(\tk)$ can be written down analytically. An approximate expression for the co-moving number density is then given by
\begin{equation}
n_{osc}a=\frac{k_{\nl}}{2\pi}\sim \frac{5}{2}\beta^{-2/5}\left(\frac{\lambda}{g}\frac{m}{2\pi}\right).
\label{eq:nosc1d}
\end{equation}
Apart from the scaling of $k$ by $\z m$, the strongest dependence is due to the $\beta$ appearing in the exponent of equation \eqref{fSol1d}. For $m/\mpl[5\times10^{-6}-5\times10^{-5}]$, $\z^{-2}[5-10]$ and $\beta[30-300]$ (equivalently $\lambda=[6.25\times 10^{-7}-6.25\times 10^{-5}]$), the above expression is consistent with the solutions of equation \eqref{nlcon1d} to within $\sim 15\%$. It is of course possible to find a better fit, however the above result serves as a useful guide for comparing with simulations. We will compare $n_{\textrm{osc}}$ obtained above with the full numerical simulations. 
  
\subsection{Numerical simulations in 1+1 dimensions}
 
In this section we follow the full non-linear evolution of the field numerically. Details of the numerical set-up and initial conditions are deferred to appendix A. First, we will review the properties of individual oscillons we expect to see emerging from our simulations. Second, we will present the numerical evolution of the field and energy density using a fiducial set of parameters and a particular realization of initial fluctuations. Third, we will average over the different realizations of the initial conditions and see how various observables evolve as a function of time and provide some statistical information about the individual characteristics of emergent oscillons. Finally, we will vary the parameters and see how the number density and fraction of energy density in oscillons depends on the parameters. We will see that the results are in good agreement (well within a factor of $2$) with our analytic estimates. 
\subsubsection{Individual oscillons}
    \begin{figure}[tbp] 
     \centering
     \includegraphics[width=5.5in]{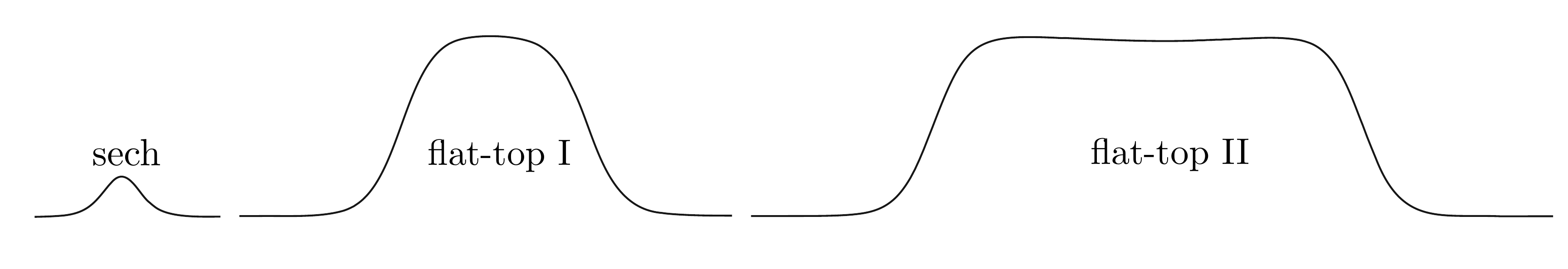} 
     \caption{The zoo of oscillon energy density profiles extracted from our simulations. At low amplitudes we see the oscillons field configuration has a ``sech" profile. The larger amplitude objects have flat tops (flat-top I). The widest objects (flat-top II) in the simulation show a slow time scale ($t\gg m^{-1}$) breathing mode. Typically their amplitudes are slightly above the critical amplitude discussed in the text. The first two types of objects are very well fit by our analytic expressions for their profiles. However, we can not analytically capture the long term breathing mode for the widest objects. }
     \label{fig:flattops}
  \end{figure}

For numerical purposes, an oscillon is defined as a persistent, localized fluctuation with a local maximum, whose energy density at this maximum is at least $5$ times the mean density. The width of an individual oscillon is defined as the size of the region where the energy density is greater that $1/e$ of its value at the center of the oscillon. The energy of an individual oscillon is defined as the energy enclosed within the above defined width of the oscillon. In, \cite{Amin:2010jq} we provided an analytic solution for the oscillon profile in an expanding universe. 
	For $(\lambda/g)^2\ll 1$, and $H\ll m$, the oscillons are described by
\begin{equation}
\varphi(t,x)=\varphi_0\sqrt{\frac{1+u}{1+u\cosh [(2\alpha\lambda/g)x]}}\cos(\omega t)
\end{equation}
where we have assumed $\lambda^2/g^2\ll1$ and have ignored terms that are higher order in $\lambda/g$.
\begin{equation}
\begin{aligned}
&u=\sqrt{1-({\alpha}/{\alpha_c})^2},\\
&\varphi_0=\sqrt{\frac{9\lambda}{10g^2}(1-u)},\\
&\omega^2=m^2\left[1-(\lambda/g)^2\alpha^2\right]
\end{aligned}
\end{equation}
This is a one parameter family of  solutions (once $\lambda,g$ and $m$ are specified), whose shape depends on $0<\alpha<\alpha_c=\sqrt{27/160}$. The width is a non-monotonic function of $\alpha$. It diverges at $\alpha\rightarrow0$ and $\alpha\rightarrow\alpha_c$. The amplitude $\varphi_0\rightarrow0$ as $\alpha\rightarrow0$, but approaches a finite value $\varphi_0\rightarrow \sqrt{9\lambda/10g^2}$ as $\alpha\rightarrow \alpha_c$. The relationship between the width and the height is shown in the left panel of figure \ref{fig:histhw}. As $\alpha\ll1$ (same as small amplitude), we get oscillons whose field profile is given by a sech function. However as $\alpha\rightarrow\alpha_c$, we get a flat-topped profile. Oscillons extracted from our simulations contain both forms of the solution (see figure \ref{fig:flattops}). Note that a ``Gaussian" profile often used in the literature is not a good fit for the flat-topped oscillons. As discussed in \cite{Farhi:2007wj,Amin:2010jq}, our solution changes character at $x^{*}\sim\alpha[(g/\lambda) H]^{-1}$ and becomes oscillatory in space due to expansion effects. The above solutions remains a good approximation as long as the width $x_e\lesssim x^{*}$, else it gets stretched out by the expansion. For fixed set of parameters $\lambda,g$ and $m$, oscillons can exist for arbitrarily large widths (with amplitudes between $0$ and $\sqrt{9\lambda/10g^2}$), but expansion effects limit their sizes. The expansion also causes a slow loss of energy from the oscillons in the form of outgoing radiation \cite{Farhi:2007wj,Amin:2010jq}. 
  \begin{figure}[tbp] 
     \centering
     \includegraphics[width=6.1in]{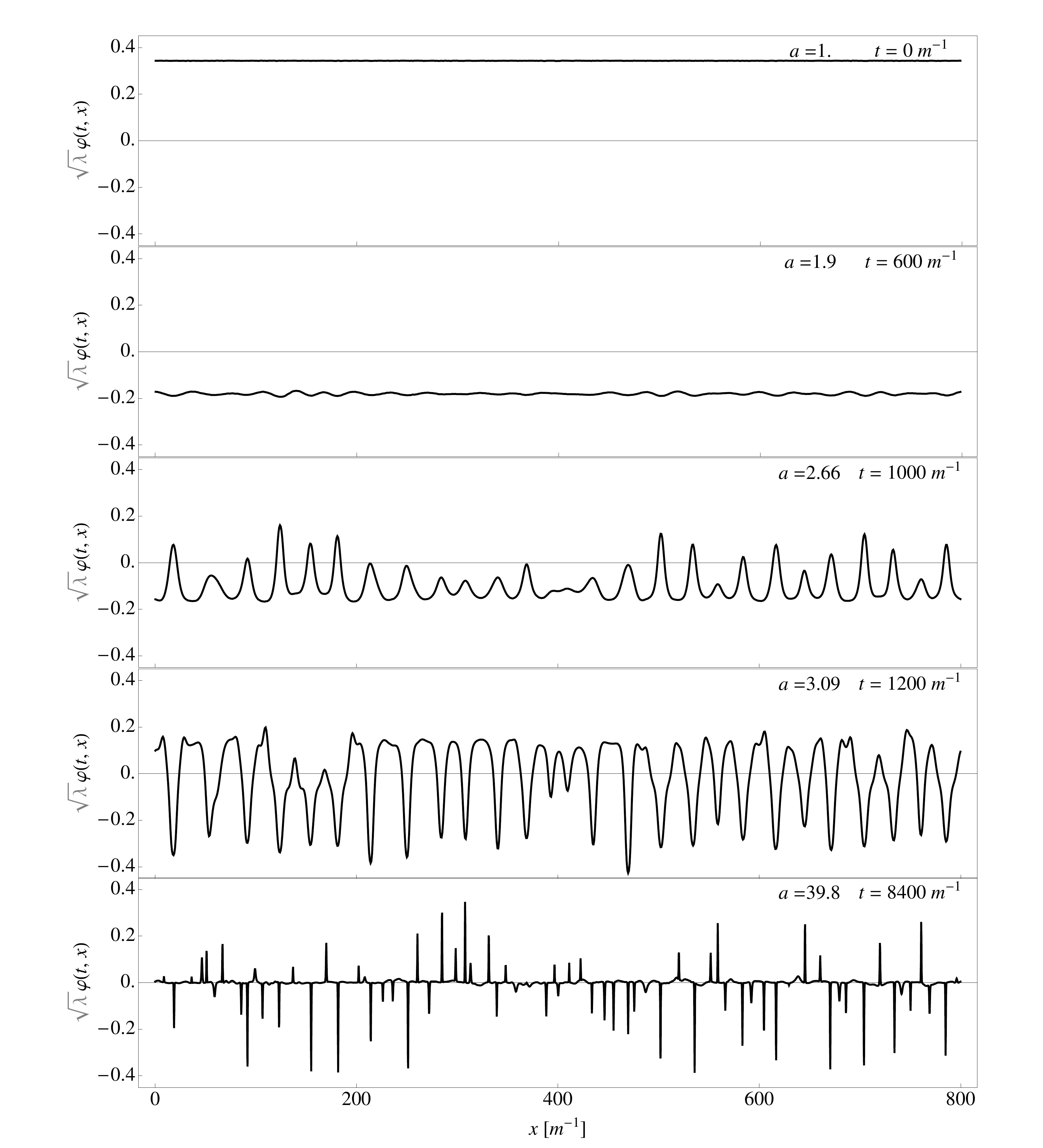} 
      \caption{The above figure shows the fragmentation of the inflaton  during the oscillatory phase of the inflaton. The large spikes in the field are oscillons. Note there characteristic scale $k_{\nl}$ at which we first develop the largest fluctuations. The number density of "first generation" oscillons is determined by this scale. Note that the oscillons appear to get thinner because their physical size is fixed as the universe expands. Here the initial Hubble parameter $H_i \approx 10^{-3}m$ whereas $\z^2=0.2$ and $m/\mpl=5\times 10^{-6}$. The co-moving size of the simulation volume is  $L\approx H_i^{-1}$ and we allow the universe to expand by $a_f=40$. An animation of the process can be found \href{http://www.mit.edu/~mamin/oscillons.html}{online}.}
    \label{fig:fgrid}
  \end{figure}
  
    \begin{figure}[tbp] 
     \centering
     \includegraphics[width=6.1in]{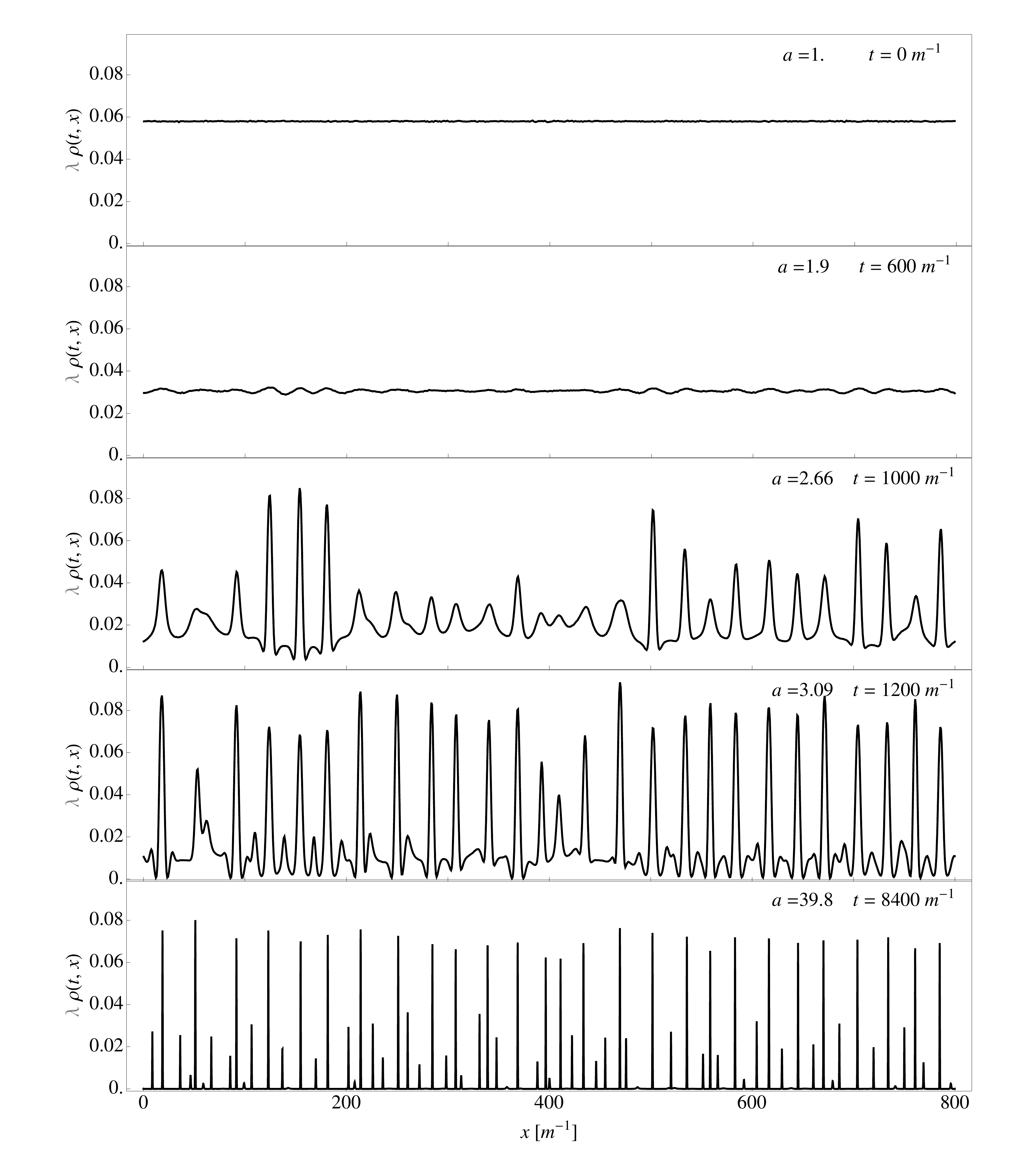} 
       \caption{The above figure shows the fragmentation of the inflaton  during the oscillatory phase of the inflaton. The large spikes in the energy density are oscillons. Note there characteristic scale $k_{\nl}$ at which we first develop the largest inhomogeneities. The number density of "first generation" oscillons is determined by this scale. Note that the oscillons appear to get thinner because their physical size is fixed as the universe expands. Here the initial Hubble parameter $H_i \approx 10^{-3}m$ whereas $\z^2=0.2$ and $m/\mpl=5\times10^{-6}$. The co-moving size of the simulation volume is  $L\approx H_i^{-1}$ and we allow the universe to expand by $a_f=40$. The energy density is expressed in units of $m^2$. An animation of the process can be found \href{http://www.mit.edu/~mamin/oscillons.html}{online}.}
     \label{fig:edgrid}
  \end{figure}

We digress briefly to consider some additional properties of extremely large width oscillons, based on the linear stability analysis of \cite{Amin:2010jq} and \cite{Hertzberg:2010yz}. In the $\alpha\ll1$ limit (``sech" like profile), $3+1$ dimensional oscillons suffer from a collapse instability when the wavelength of the perturbations are comparable to the width of the oscillons \cite{Amin:2010jq}. As $\alpha$ increases, moving towards the flat-topped configurations, this instability disappears. In $1+1$-dimensions,  the collapse instability is not present. In the $\alpha\rightarrow\alpha_c$ limit, extremely wide oscillons can efficiently transfer energy to $(k\sim \sqrt{3}m)$ perturbations \cite{Hertzberg:2010yz}. As a result, we expect an upper limit on the width of oscillons seen in our simulations. 
We also note that we see some breathing-mode configurations of localized energy densities when $\alpha\gtrsim \alpha_c$, for whom we do not have an analytic description (see flat-top II in figure \ref{fig:flattops}). We conjecture that they are likely to be bound states of two oscillons.

\subsubsection{Field and energy density evolution: single realization}
   \label{sec:nosc1d}
Let us now follow the evolution of the field and energy densities for a single run. In figures \ref{fig:fgrid} and  \ref{fig:edgrid} we show the evolution of the field and the energy density. The values of the parameters were chosen to be $m/\mpl=5\times 10^{-6}$, $(\lambda/g)^2=0.2$ and $\lambda\approx 3\times 10^{-7}$ ($\beta=50$). The large spikes in the field and energy density are oscillons. Note that the oscillons appear to get thinner because their physical size is fixed as the universe expands and the horizontal axes are labelled in co-moving co-ordinates. The initial box size is $L\approx H_i^{-1}=800 m^{-1}$. At the end $L_f\approx 6.4 H_f^{-1}.$ In the simulations $H\approx H_i/\sqrt{a}$.

For $a\ll a_{\nl}\approx 2.7$ we do not see any significant amplification. At $a\sim a_{\nl}$ significant deviations from the homogeneous energy density appear with a characteristic wavenumber $k_{\nl}$. This leads to the formation of the first generation of large oscillons. Their number density is predicted by the formula in equation \eqref{eq:nosc1d}. A second burst is seen at $a_2$, with energies smaller than the first generation ones. There could be subsequent bursts, however the energies of the oscillons produced tend to be significantly lower than the first burst. In addition, these low energy oscillons (with small amplitudes) have large widths and get stretched out by the Hubble expansion. 

\subsubsection{Field and energy density evolution: statistics}

We ran 10 simulations with the same parameters, with different initial conditions for the fluctuations drawn from a Gaussian distribution (also see appendix A). In figure \ref{fig:histhw} (right panel) we show a histogram of oscillon energies at $a_f=40$. The distribution of energy densities is bimodal at late times and corresponds to the two generations of oscillons. Based on this histogram we call oscillons in the right lobe of the bimodal distribution, the ``first" generation oscillons. One might wonder if the lumps we are seeing are indeed oscillons. We can get rid of of the false positives by comparing the height and width of the energy density of the lumps with the theoretical relationship (see the left panel in figure \ref{fig:histhw}). As seen in figure \ref{fig:edgrid}, at late times, all large lumps that remain are oscillons. Again, note the bimodal distribution of widths and heights of the oscillons corresponding to the two generations. 

      \begin{figure}[tbp] 
     \centering
     \includegraphics[width=5.5in]{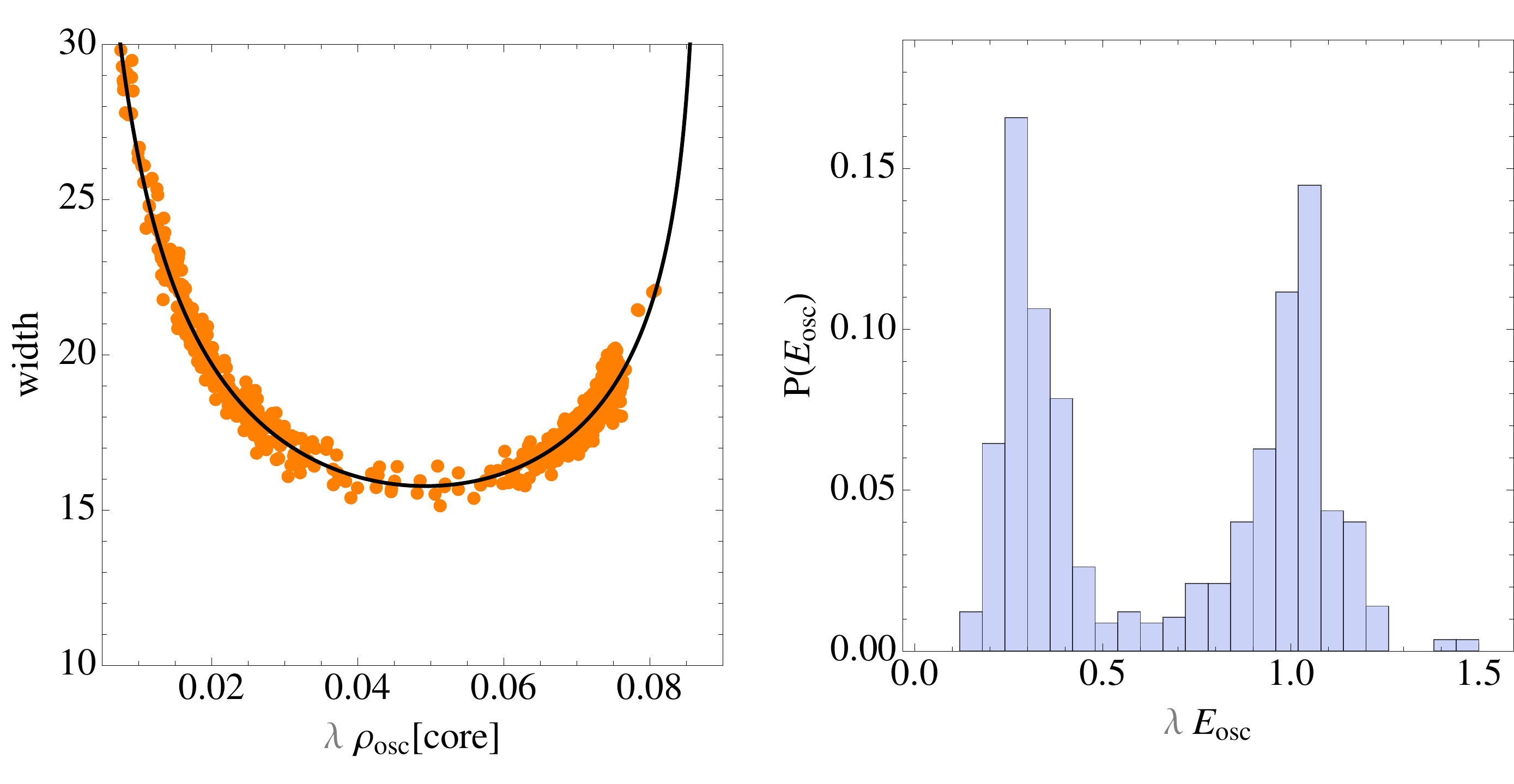} 
     \caption{[$1+1$-dimensions] In the left panel, we show the non-monotonic relationship between the width and core amplitude of the oscillon energy densities based on our analytic solution (solid black line). The orange points are from the energy density configurations flagged as oscillons in our simulations. At large core amplitude, we approach a flat-topped profile. In the right panel, a histogram of the energies of oscillons is shown [$\lambda\approx 3\times10^{-7}, \z^2=0.2, m/\mpl=5\times 10^{-6}$ with 10 different initial condition realizations].  Two distinct oscillon populations are clearly visible, with the larger energy ones appear first in the simulations. 
All variables are expressed in units of appropriate powers of mass $m$. An animation of the process can be found \href{http://www.mit.edu/~mamin/oscillons.html}{online}.}
     \label{fig:histhw}
  \end{figure}

\begin{figure}[tbp] 
     \centering
     \includegraphics[width=5.5in]{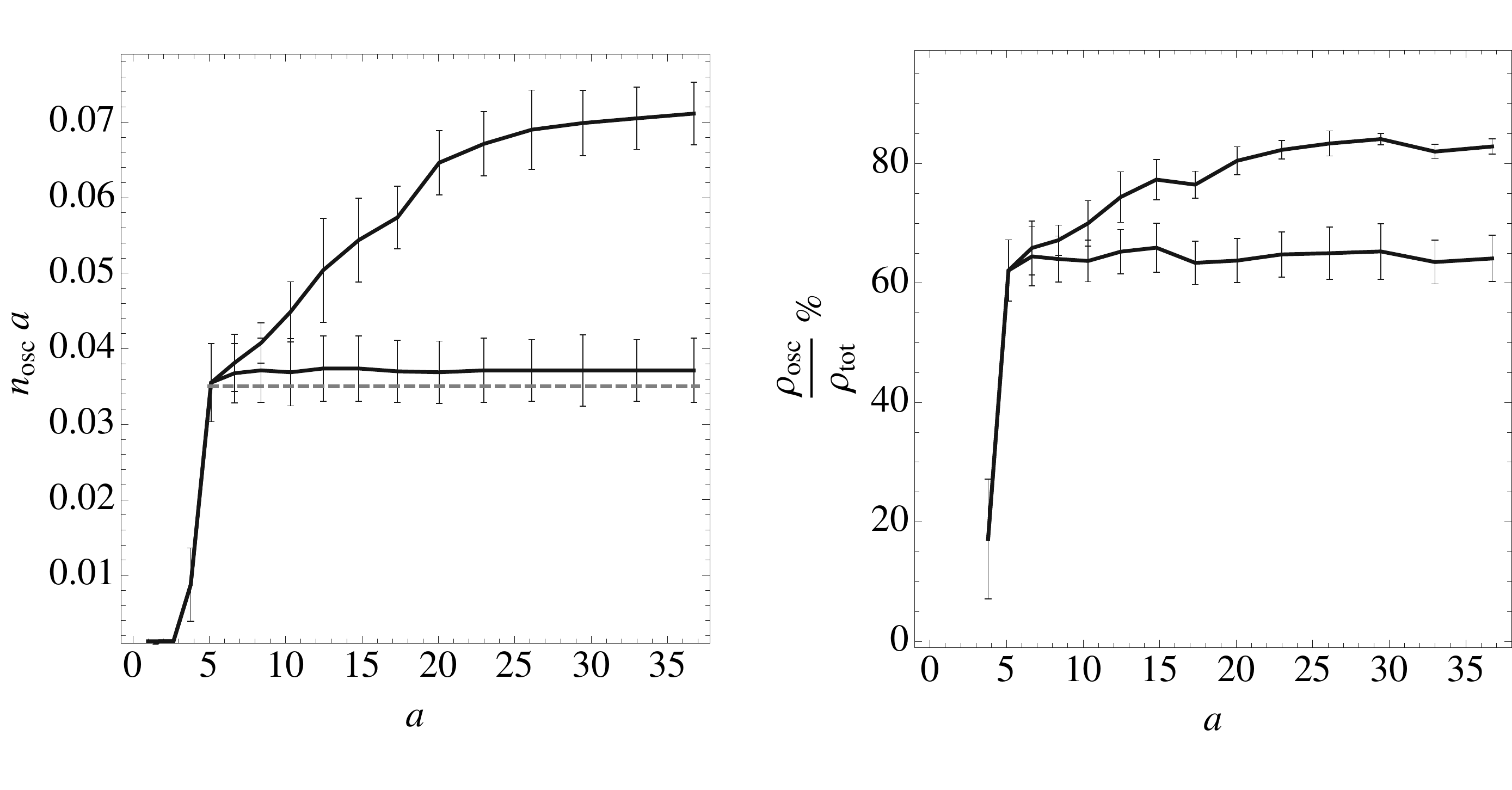} 
     \caption{[$1+1$-dimensions] The left panel shows the number density of oscillons (total and first generation) as a function of the scalefactor. Note that they are produced at $a_{\nl}\sim 2.7$ and thereafter freeze out. The dashed line is our analytic estimate in \eqref{eq:nosc1d}. The resulting energy density in oscillons after averaging the runs over 10 realizations of the initial conditions is shown on the left. The 2 sigma error bars quantify the variations between different runs. Note that most of the energy density is in the first generation of oscillons. The values of the parameters used were $\lambda\approx 3\times10^{-7}$ (or $\beta=50$), $\z^2=0.2$ and $m/\mpl=5\times 10^{-6}$.  Number density is expressed in units of the mass $m$.}
     \label{fig:enosca}
  \end{figure}

In figure \ref{fig:enosca} (right panel) we show the number density of oscillons (total and first generation) as a function of time. Large oscillons are produced at $\sim a_{\nl}$ and thereafter freeze out with the expansion. The dashed line is our analytic estimate in \eqref{eq:nosc1d}. The results of fraction of energy density in oscillons (after averaging the runs over 10 realizations of the initial conditions) is shown in figure \ref{fig:enosca} (left panel). The 2 sigma error bars quantify the variations between different runs. Note that most of the energy density is in the first generation of oscillons. The values of the parameters used were $\lambda\approx 3\times10^{-7}$ (or $\beta=50$), $\z^2=0.2$ and $m/\mpl=5\times 10^{-6}$.

\subsubsection{Field and energy density evolution: parameter dependence}
We will now vary the parameters of the model and see if our results match the semianalytic estimate. 
Figure \ref{fig:nlfirst1d} (b) shows the final number density of oscillons at $a_f=40$ as a function of $\beta$ for $\z^2=0.1$(black) and $\z^2=0.2$ (orange). The dashed curve  is our analytic estimate. It is somewhat remarkable, that in spite of the extremely non-linear dynamics of oscillon formation, our analysis gets the number density to well within a factor of $2$ and correctly captures the variation with parameters.

At small $\beta\lesssim 20$, parametric resonance stops being effective and we get fewer and smaller (in amplitude) oscillons. At large $\beta$ we notice a systematic deviation from our analytic estimate. As seen in figure \ref{fig:nlfirst1d}(b), the analytic result over-estimates the number of oscillons at large $\beta$. However, at large $\beta$, the rate of expansion is slow. As a result, in our simulations we get a large number of oscillon-oscillon interactions, which leads to some of them being disrupted. This was not taken into account in our estimate. Oscillons can merge, scatter off each other or be disrupted, depending on their amplitudes and phases. Our exploratory investigation with oscillon collisions indicates that the outcome of the collision depends on the phase, amplitude as well as shape of the oscillons (also see \cite{Hindmarsh:2007jb}). The collisions range between almost completely elastic collisions to highly inelastic ones (though we are almost always left with one or more oscillons). The details of oscillon-oscillon interaction or interactions with other large non-linearities is not completely understood and certainly warrants further investigation. 

A second effect is that as $k_{\nl}$ gets smaller, we get very wide oscillons. These objects suffer from a Floquet instability at $k\sim \sqrt{3}m$ \cite{Amin:2010jq, Hertzberg:2010yz} and can potentially destabilize the oscillons. We also note that in our simulations, the largest width objects (see flat-top II in figure \ref{fig:flattops}) have an energy density that stays localized but exhibits a long time scale ($t\gg m^{-1}$), breathing mode which do not fully understand. 

\begin{figure}[tbp] 
     \centering
     \includegraphics[width=5.5in]{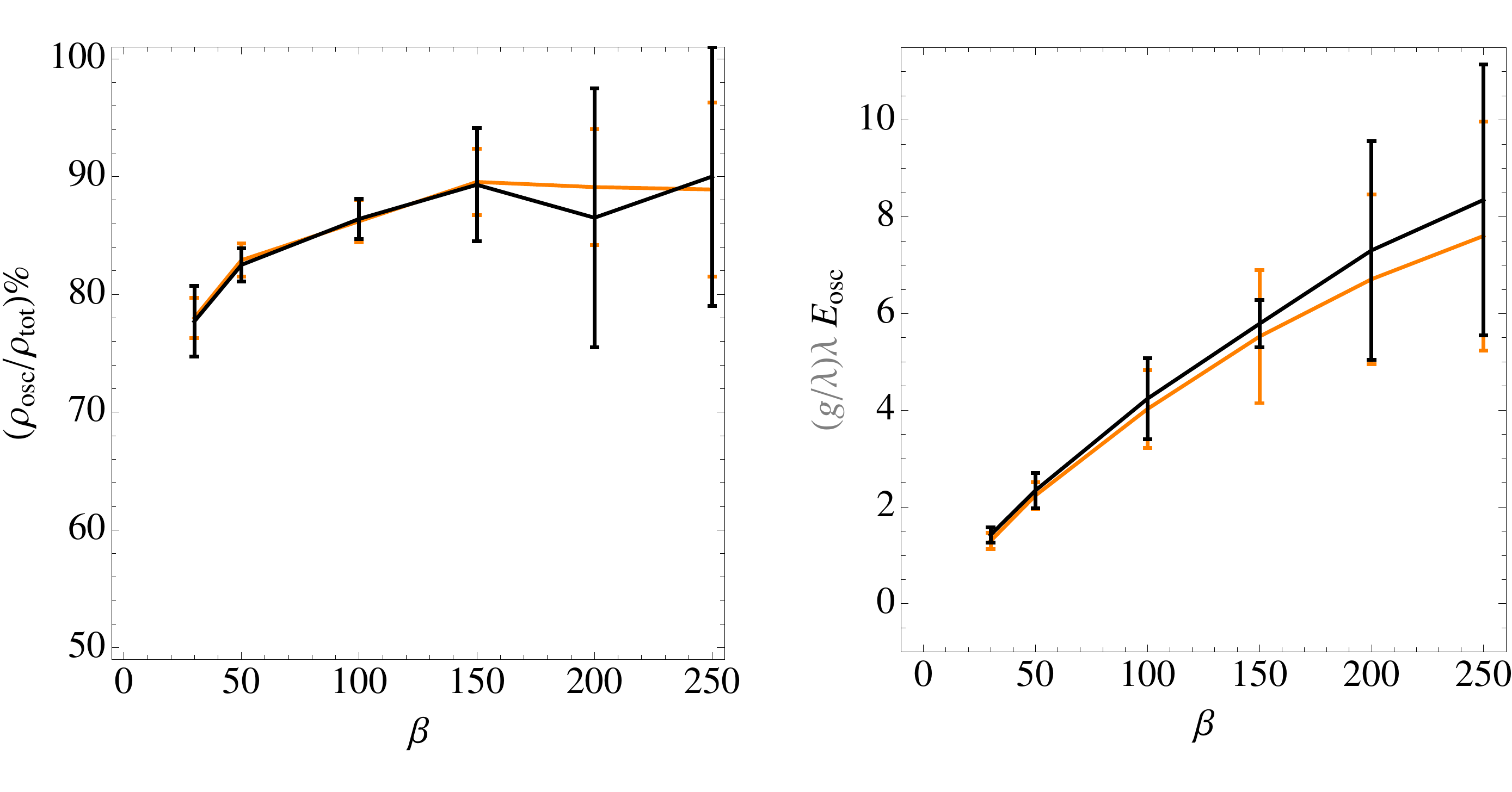} 
     \caption{[$1+1$-dimensions] The left panel shows the percentage of energy density in oscillons at the end of our simulations ($a_f=40$) for different values of $\beta=\sqrt{\lambda}\z(\mpl/m)$. The orange and black curves with 2 sigma error bars correspond to simulations with parameters $\z^2=0.1$ and $0.2$ respectively. Note that for our fiducial value of $m/\mpl=5\times 10^{-6}$, we have $\lambda\sim 10^{-10}\beta^2$. The fraction of energy density in oscillons decreases at low $\beta$ (large $H$) because resonance is not efficient enough to significantly amplify initial fluctuations. The right panel shows the mean energy of the first generation oscillons. Note that it increases with increasing $\beta$ (decreasing $H$). This is consistent with the idea that at large $\beta$ oscillons form from longer wavelength fluctuations, thus yielding fewer, larger oscillons. Energy is expressed in units of the mass $m$.}
     \label{fig:efracE}
  \end{figure}

The oscillons tend to dominate the energy density of the field. In figure \ref{fig:efracE} (right), we quantify this statement. As expected, at very small $\beta$ we cannot generate oscillons efficiently, so this fraction must increase as a function of $\beta$. For the parameter range considered, oscillons take up more that $75-90\%$ of the energy density of the field. Note that for numerical purposes we defined the oscillon energy as the energy contained within a width where the energy density is above $1/e$ of its core value. Thus, we we are ignoring the energy density in the tails. 

The energy of typical oscillons (at a fixed $\lambda/g$) increases with increasing $\beta$. That is, as the Hubble parameter gets smaller, we get larger, more massive oscillons without a significant change in the fraction of energy density in oscillons. The mean energy of first generation oscillons is shown in figure \ref{fig:efracE} (left). The two colors represent different values of $\z^2$[0.1,0.2]. The mean and two sigma error bars are obtained from a sample containing a few thousand oscillons (100 realizations, 10 for each value of $\beta$) . 

\section{Discussion}
\label{sec:disc}
In this paper we have investigated the post inflationary, emergence of oscillons in a class of single field inflaton models. We gave analytic results for $3+1$ and $1+1$-dimensional cases and numerical results for the $1+1$-dimensional case. Starting from zero point fluctuations of the inflaton during its oscillatory regime, we provided an (approximate) analytic description of the linear evolution of the fluctuations and provided a condition necessary for significant amplification of the fluctuations. Using this linear analysis we calculated the characteristic scale which is the first to become non-linear. We hypothesized that it is this scale that determines the number density of oscillons. We checked our analysis in detail with $1+1$ dimensional simulations in an expanding universe, varying different parameters over an order of magnitude. Here we found agreement between our analytical and numerical results to well within a factor of $\sim 2$. The number density decreased and the size of oscillons increased with decreasing Hubble (all other parameters fixed). A detailed analysis revealed that the individual characteristics of the oscillons (in particular the natural emergence of flat-top oscillons) extracted from our simulations were in excellent agreement with the analytic results of our previous paper \cite{Amin:2010jq}. We found that the fraction of energy density in oscillons as the parameters were varied was $75\%-90\%$. We also pointed out some interesting phenomenon seen in the simulations which we cannot completely account for quantitatively. These included the production of oscillons in more than one burst (however, see \cite{Gleiser:2003uu}), oscillon-oscillon interactions at slow Hubble rates and the slow breathing modes of extremely wide, flat-top oscillons. 

There are many ways in which our analysis could be extended. In particular, we ignored the effects of higher order resonance bands. Although reasonable for the model under consideration, this need not be true for other models and should be included in the analysis. The effect of coupling to other fields needs to be investigated \cite{Hertzberg:2010yz}. A more careful analysis of the initial conditions and the evolution of fluctuations including gravitational perturbations is also needed. Most importantly, our numerical analysis was done in $1+1$-dimensions. In an upcoming publication \cite{Amin:2010}, we will provide detailed numerical results for the $3+1$ dimensional case and compare it to the analytical estimates provided in this paper. 

In summary, to understand the cosmological consequences of oscillons, it is important to have a prediction for their number densities as well as their individual characteristics. For the model under consideration, we provided both in terms of the parameters in the inflaton Lagrangian. The techniques developed here should be directly applicable in a broader class of models.

 \section{Acknowledgements}
We would like to thank Nabil Iqbal for discussions regarding the $1+1$ dimensional simulations, Matt Johnson and David Shirokoff for help regarding Floquet theory, Raphael Flaugher and Richard Easther for the discussions on the possible role of gravity in determining initial conditions, David Gosset and Alan Guth for insights regarding the zero point initial conditions and Mark Hertzberg regarding the stability of oscillons to short wavelength perturbations. We would also like to thank  Ed Bertschinger, Eddie Farhi,  Evangelos Sfakianakis, Surjeet Rajendran, Hal Finkel and Ruben Rosales for many stimulating discussions. We would especially like to thank David Shirokoff and Richard Easther for a careful reading of the manuscript and useful suggestions for its improvement. We acknowledge the support from a Pappalardo Fellowship at MIT.

 \section*{Appendix A: Numerical set-up and initial conditions}
In this appendix, we outline our numerical set-up for the $1+1$ dimensional simulations.  For numerical purposes,  we find it convenient to work with dimensionless spacetime variables $mx^{\mu}$ as well as the scaled field $\varphi_{\textrm{p}}=\lambda^{1/2}\varphi$. We will work with conformal time $d\eta=a^{-1}dt$. Under these changes, the equations of motion now become:

 \begin{equation}
 \begin{aligned}
 & \varphi_\p''-\partial_x^2\varphi_\p+a^2(\eta)V_\p'(\varphi_\p)=0\\
  &V_\p= \frac{1}{2}\varphi_\p^2-\frac{1}{4}\varphi_\p^4+\frac{1}{6\zeta^{2}}\varphi_\p^6
  \end{aligned}
  \end{equation}
 where the `prime' stands for derivatives with respect to conformal time and $\zeta=\lambda/g$. To avoid clutter, we will drop the $\textrm{p}$ subscript. Using conformal time is particularly convenient for numerical purposes since it gets rid of the linear derivative term. 
  
  We will discretize the above equations in space and conformal time. We will denote $\varphi(x,\eta)=\varphi(i\Delta x,j\Delta \eta)\equiv \varphi_{i,j}$ and $a(\eta)=a(j\Delta\eta)\equiv a_j$. The symmetric space and time derivatives become:
 
 \begin{equation}
 \begin{aligned}
 \varphi'_{i,j}&=\frac{\varphi_{i,j+1}-\varphi_{i,j-1}}{2\Delta\eta},\\
\partial_x\varphi_{i,j}&=\frac{\varphi_{i+1,j}-\varphi_{i-1,j}}{2\Delta x},\\
\varphi''_{i,j}&=\frac{\varphi_{i,j+1}-2\varphi_{i,j}+\varphi_{i,j-1}}{\Delta\eta^2},\\
\partial_x^2\varphi_{i,j}&=\frac{\varphi_{i+1,j}-2\varphi_{i,j}+\varphi_{i-1,j}}{\Delta x^2}.\\
 \end{aligned}
 \end{equation}
 The evolution equations are given by:
 \begin{equation}
 \begin{aligned}
 \varphi_{i,j+1}=2\varphi_{i,j}-\varphi_{i,j-1}+s^2\left[\varphi_{i-1,j}+\varphi_{i+1,j}-2\varphi_{i,j}\right]
 -(\Delta\eta)^2a_j^2V'(\varphi_{i,j}),
 \end{aligned}
 \end{equation}
 where $s=\Delta\eta/\Delta x$. To evolve the system forward in time we need $\varphi_{i,0}$ and $\varphi_{i,1}$. $\varphi_{i,0}$ is the initial value of the field $\varphi_{i,0}=\varphi(0,i\Delta x)$ whereas $\varphi_{i,1}$ be constructed from $\varphi_{i,0}$ and $\varphi'_{i,0}=\varphi'(0,i\Delta x)$ as follows
 \begin{equation}
 \begin{aligned}
 \varphi_{i,1}=\varphi_{i,0}+\varphi'_{i,0}\Delta\eta+\frac{s^2}{2}(\varphi_{i+1,0}-2\varphi_{i,0}+\varphi_{i-1,0})-\Delta\eta^2V'(\varphi_{i,0})
 \end{aligned}
 \end{equation}
 We employ periodic boundary conditions. The initial field value and it's derivative are specified using (see \cite{Farhi:2007wj}):
 \begin{equation}
 \begin{aligned}
 \varphi_{i,0}=\bp_{i,0}+\frac{1}{\sqrt{L}}\sum_{-N/2+1}^{N/2}\sqrt{\frac{1}{2\omega_n}}\left[\alpha_ne^{\iota k_n (i\Delta x)}+c.c\right],\\
  \varphi'_{i,0}=\frac{1}{\sqrt{L}}\sum_{-N/2+1}^{N/2}\frac{1}{\iota}\sqrt{\frac{\omega_n}{2}}\left[\alpha_ne^{\iota k_n (i\Delta x)}-c.c\right],\
  \end{aligned}
 \end{equation} 
 where $k_n=2\pi L^{-1}n$ and $\omega_n=\sqrt{1+\left(2\sin \frac{k_n\Delta x}{2}/{\Delta x}\right)^2}$.
 Here, $\alpha_n$ are complex numbers whose phases drawn at random from $(0,2\pi]$ whereas their amplitudes are drawn from the Gaussian distribution with variance $\langle|\alpha_n|^2\rangle=\lambda/2$, consistent with the description of a massive field in it's ground state. A number of comments are in order. We have expressed $k_n$ and $\omega_n$ in units of $m$. We have ignored the interaction terms in specification of the ground state. We do not know of any prescription where these can be taken into account for the zero point fluctuations. The appearance of $\lambda$ in the amplitude of the fluctuation is somewhat misleading since we did not take into account the interaction terms. Here it appears simply because we are working with the scaled version of the field $\varphi_{\p}=\sqrt{\lambda}\varphi$. We chose to treat the evolution classically, with the anticipation that the occupation number per mode will grow rapidly as they undergo parametric resonance (see for example \cite{Khlebnikov:1996mc}). 
 
 In addition, we have specified the initial conditions in Minkowski space. This is reasonable since, the scales of interest are much smaller than $H^{-1}$. However, this significantly underestimates the fluctuations close to the scale of the Horizon. The nature of the spectrum there depends on the details of the evolution of the field before $t_i$ and the shape of the potential beyond $\bp_i$. Since we have not specified these in this paper, we take this conservative prescription of the initial conditions as our starting point. Another approximation is present at $k\gg m$. As modes redshift we need to continuously re-populate the high $k$ modes on the lattice. Since these high $k$ modes do not undergo efficient parametric resonance (based on the structure of the Floquet diagram), we do not re-populate these modes. One has to be somewhat cautious here since some oscillons have discrete narrow band instabilities in the $k\gg m$ region \cite{Amin:2010jq, Hertzberg:2010yz}. This will not be captured by the simulation. 
 
 Now, we need to choose the spatial and temporal resolution. Oscillons maintain a fixed physical size as the universe expands. This means that we need to improve our spatial resolution as the universe expands. We do so by doubling the grid points every time the universe expands by a factor of 2. We make sure $\Delta \eta <\Delta x$ is always satisfied, by refining the time step along with the spatial resolution. The approach is similar to the one used in \cite{Farhi:2007wj}. We interpolate the field between adjacent grid points to improve the spatial resolution, but use the equations of motion themselves (with assymetric time steps) for improving the temporal resolution. We need to make sure that we capture the fastest spatial as well as temporal oscillations of the system, both of which should be (at least) smaller that the inverse mass scale of the problem. We start with an initial grid spacing of $\Delta x_0=0.5$. The initial time step: $\Delta \eta_0=0.2\Delta x_0$.  By construction we cannot resolve spatial structures much smaller than $0.5$. The size of the initial box is $L_0=800$. We allow the simulation to run till $a_f=40$. Our overall energy conservation (including the effects of expansion) is at the level of one part in $10^3$. Almost all field configurations that are flagged as oscillons are resolved with $>20$ grid points. We have varied the spatial and temporal resolution to make sure that no significant qualitative differences are seen. Nevertheless, the final positions of the oscillons do shift as the resolution is varied. 

To evolve $a(\eta)$ we need $H$ at each time step: $a_{j+1}=a_j\left(1+a_jH\Delta\eta\right)$. This can be obtained using the Friedmann equation $H=1/(3\mpl^2)\langle\rho\rangle$ where $\langle\rho\rangle$ stands for the spatially averaged energy density.  In practice, we find that $H\propto a^{-1/2}$ reproduces the expansion history to a percent level accuracy. As mentioned in the main body of the text, in $1+1$-dimensions this amounts to a prescription for the evolution of the background since in $1+1$-dimensions, the Einstein tensor is identically zero.

 \section*{Appendix B: Floquet exponent}
 In this section we will derive the Floquet exponent in equation \eqref{eq:nxfloquet}. The approach is similar to the one adopted in \cite{Landau:1976} for a parametrically excited harmonic oscillator\footnote{We thank Raphael Flaugher for pointing us to this reference.}. 
We start with the equation of motion for the homogeneous field and linear fluctuations around the homogeneous solution:
\begin{equation}
\begin{aligned}
&\partial_t^2\bp+m^2\bp-\lambda\bp^3+\frac{g^2}{m^2}{\bp^5}=0,\\
&\partial_t^2\delta\varphi-\nabla^2 \delta\varphi +\left[m^2-3\lambda\bp^3+\frac{5g^2}{m^2}{\bp^4}\right]\delta\varphi=0.\\
\end{aligned}
\end{equation}
It is convenient to define a dimensionless, scaled version of the field
\begin{equation}
\phi=\sqrt{\lambda}\frac{\varphi}{m},
\end{equation}
and dimensionless spacetime variables $x^\mu\rightarrow mx^\mu$. In terms of these variables we have
\begin{equation}
\begin{aligned}
&\partial_t^2\sbp+\sbp-\sbp^3+\left(\frac{g}{\lambda}\right)^2\sbp^5=0,\\
&\partial_t^2 \delta\phi +\left[-\nabla^2 +1-3\sbp^2+5\left(\frac{g}{\lambda}\right)^{\!2}\sbp^4\right] \delta\phi  = 0.
\end{aligned}
\end{equation}
We will be working under the assumption $\epsilon^2\equiv(\lambda/g)^2\ll 1$. Under this assumption, the homogeneous background equation has a solution of the form:
\begin{equation}
\sbp(t)\approx \epsilon\Phi_0 \cos\omega t+\mathcal{O}[\epsilon^3],
\end{equation}
where $\Phi_0$ can be of order unity. The frequency of oscillation is
\begin{equation}
\omega\approx 1-\epsilon^2\left(\frac{3}{8}\Phi_0^2-\frac{5}{16}\Phi_0^4\right)+\mathcal{O}[\epsilon^4].
\end{equation}
The equation of motion for the perturbation $\delta\phi$ (in Fourier space) is 
\begin{equation}
\label{eq:pertFourier}
  \partial_t^2 \delta\phi +\left[1+k^2- \epsilon^2\left( 3\Phi_0^2\cos^2\omega t- 5\Phi_0^4\cos^4 \omega t\right) \right] \delta\phi  = 0.
\end{equation}
We can rewrite the above equation as:
\begin{equation}
\label{eq:pertFourier}
  \partial_t^2 \delta\phi +\left[\Omega_k^2+\epsilon^2\left(\beta \cos 2\omega t+\gamma\cos 4\omega t\right) \right] \delta\phi  = 0,
\end{equation}
where 
\begin{equation}
\begin{aligned}
 &       \Omega_k^2 = 1+k^2-\epsilon^2\left(\frac{3}{2}\Phi_0^2 + \frac{15}{8}\Phi_0^4\right), \\
 &    \beta = -\frac{3}{2}\Phi_0^2 + \frac{5}{2}\Phi_0^4, \\
 &    \gamma = \frac{5}{8}\Phi_0^4.
        \end{aligned}
        \end{equation}
We shall look for solutions of the form
\begin{equation}
\delta\phi(t)=\sum_{n=1,3\hdots} \left[a_n(t)\cos n\Omega_k t+b_n(t)\sin  \cos n\Omega_k t\right]
\end{equation}
where $a_n$ and $b_n$ are slowly varying compared to the oscillatory terms. The even $n$ terms are decoupled from the odd-terms and can be set to zero. Plugging this form of the solution into equation \eqref{eq:pertFourier}, dropping the second time derivatives ($\ddot{a}_n$ and $\ddot{b}_n$) and only keeping terms up to order $\epsilon^2$, we get:

\begin{equation}
\left( \begin{array}{c}
\dot{a}_1  \\
\dot{b}_1  \\
\end{array} \right)=\frac{1}{4\omega}\left( \begin{array}{cc}
0 & \beta+2(\omega^2-\Omega_k^2)  \\
-\beta+2(\omega^2-\Omega_k^2)  & 0  \\
\end{array} \right)\left( \begin{array}{c}
a_1  \\
b_1  \\
\end{array} \right)+\mathcal{O}[\epsilon^2]
\end{equation}

The system can be easily diagonalized to obtain the following solution, conveniently expressed in its Eigen-basis:
\begin{equation}
\left( \begin{array}{c}
a_1(t)  \\
b_1(t)  \\
\end{array} \right)=
c_1e^{\mu_k t}\left( \begin{array}{c}
-1  \\
4\omega \mu_k  \\
\end{array} \right)+
c_2e^{-\mu_k t}\left( \begin{array}{c}
1  \\
4\omega \mu_k  \\
\end{array} \right)+\mathcal{O}[\epsilon^2],
\end{equation}
where the eigenvalue
\begin{equation}
\mu_k=\frac{1}{4\omega}\sqrt{\beta^2-4(\omega^2-\Omega_k^2)^2}
\end{equation}
is the desired Floquet exponent. Note that after an initial transient, $\delta\phi\propto e^{{\mu_k}t}$ (ignoring the oscillatory piece). The higher harmonic terms are higher order in $\epsilon$.

To lowest order in $\epsilon$, the Floquet exponent is given by
\begin{equation}
\mu_k=\frac{k}{2}\sqrt{\frac{3}{2}\epsilon^2{\Phi_0}^2\left(1-\frac{5}{3}\Phi_0^2\right)-k^2}.
\end{equation}
Reverting back to the original, unscaled variables used in the main body of the text ($\bp=\left(m\epsilon\Phi_0/\sqrt{\lambda}\right)\cos\omega t$ and $k\rightarrow k/m$), we have
\begin{equation}
\mu_k=\frac{k}{2}\sqrt{\frac{3\lambda}{2}\left(\frac{\bp}{m}\right)^{\!2}\left(1-\frac{\bp^2}{\bp_i^2}\right)-\left(\frac{k}{m}\right)^{\!2}},
\end{equation}
where $\bp_i=\sqrt{3\lambda/5g^2}m$.

\end{document}